\documentclass[twocolumn]{IEEEtran}
\usepackage{amsmath,amssymb,bm,cite,color}
\usepackage{theorem}
\usepackage{multirow}
\usepackage{graphicx}
\usepackage{subfigure}
\usepackage{tikz}
\usetikzlibrary{decorations.pathreplacing}
\usepackage{enumerate}
\usepackage[linesnumbered,ruled]{algorithm2e}
\usepackage{hyperref}
\usepackage{cleveref}

\newtheorem{theorem}{Theorem}[section]
\newtheorem{lemma}{Lemma}[section]

\newtheorem{fact}{Fact}[section]

\newtheorem{corollary}{Corollary}[section]
\newtheorem{definition}{Definition}[section]

\newtheorem{observation}{Observation}[section]
\newtheorem{example}{Example}[section]
\newtheorem{remark}{Remark}[section]

\newcommand\nc\newcommand
\nc{\cA}{\mathcal{A}}\nc{\cB}{\mathcal{B}}\nc{\cC}{\mathcal{C}}\nc{\cD}{\mathcal{D}}
\nc{\cE}{\mathcal{E}}\nc{\cF}{\mathcal{F}}\nc{\cG}{\mathcal{G}}\nc{\cH}{\mathcal{H}}
\nc{\cI}{\mathcal{I}}\nc{\cJ}{\mathcal{J}}\nc{\cK}{\mathcal{K}}\nc{\cL}{\mathcal{L}}
\nc{\cM}{\mathcal{M}}\nc{\cN}{\mathcal{N}}\nc{\cO}{\mathcal{O}}\nc{\cP}{\mathcal{P}}
\nc{\cQ}{\mathcal{Q}}\nc{\cR}{\mathcal{R}}\nc{\cS}{\mathcal{S}}\nc{\cT}{\mathcal{T}}
\nc{\cU}{\mathcal{U}}\nc{\cV}{\mathcal{V}}\nc{\cW}{\mathcal{W}}\nc{\cX}{\mathcal{X}}
\nc{\cY}{\mathcal{Y}}\nc{\cZ}{\mathcal{Z}}

\nc{\bba}{\mathbf{a}}\nc{\bbb}{\mathbf{b}}\nc{\bbc}{\mathbf{c}}\nc{\bbd}{\mathbf{d}}
\nc{\bbe}{\mathbf{e}}\nc{\bbf}{\mathbf{f}}\nc{\bbg}{\mathbf{g}}\nc{\bbh}{\mathbf{h}}
\nc{\bbi}{\mathbf{i}}\nc{\bbj}{\mathbf{j}}\nc{\bbk}{\mathbf{k}}\nc{\bbl}{\mathbf{l}}
\nc{\bbm}{\mathbf{m}}\nc{\bbn}{\mathbf{n}}\nc{\bbo}{\mathbf{o}}\nc{\bbp}{\mathbf{p}}
\nc{\bbq}{\mathbf{q}}\nc{\bbr}{\mathbf{r}}\nc{\bbs}{\mathbf{s}}\nc{\bbt}{\mathbf{t}}
\nc{\bbu}{\mathbf{u}}\nc{\bbv}{\mathbf{v}}\nc{\bbw}{\mathbf{w}}\nc{\bfx}{\mathbf{x}}
\nc{\bby}{\mathbf{y}}\nc{\bbz}{\mathbf{z}}

\nc{\bbA}{\mathbf{A}}\nc{\bbB}{\mathbf{B}}\nc{\bbC}{\mathbf{C}}\nc{\bbD}{\mathbf{D}}
\nc{\bbE}{\mathbf{E}}\nc{\bbF}{\mathbf{F}}\nc{\bbG}{\mathbf{G}}\nc{\bbH}{\mathbf{H}}
\nc{\bbI}{\mathbf{I}}\nc{\bbJ}{\mathbf{J}}\nc{\bbK}{\mathbf{K}}\nc{\bbL}{\mathbf{L}}
\nc{\bbM}{\mathbf{M}}\nc{\bbN}{\mathbf{N}}\nc{\bbO}{\mathbf{O}}\nc{\bbP}{\mathbf{P}}
\nc{\bbQ}{\mathbf{Q}}\nc{\bbR}{\mathbf{R}}\nc{\bbS}{\mathbf{S}}\nc{\bbT}{\mathbf{T}}
\nc{\bbU}{\mathbf{U}}\nc{\bbV}{\mathbf{V}}\nc{\bbW}{\mathbf{W}}\nc{\bfX}{\mathbf{X}}
\nc{\bbY}{\mathbf{Y}}\nc{\bbZ}{\mathbf{Z}}

\nc{\sA}{\mathsf{A}}\nc{\sB}{\mathsf{B}}\nc{\sC}{\mathsf{C}}\nc{\sD}{\mathsf{D}}
\nc{\sE}{\mathsf{E}}\nc{\sF}{\mathsf{F}}\nc{\sG}{\mathsf{G}}\nc{\sH}{\mathsf{H}}
\nc{\sI}{\mathsf{I}}\nc{\sJ}{\mathsf{J}}\nc{\sK}{\mathsf{K}}\nc{\sL}{\mathsf{L}}
\nc{\sM}{\mathsf{M}}\nc{\sN}{\mathsf{N}}\nc{\sO}{\mathsf{O}}\nc{\sP}{\mathsf{P}}
\nc{\sQ}{\mathsf{Q}}\nc{\sR}{\mathsf{R}}\nc{\sS}{\mathsf{S}}\nc{\sT}{\mathsf{T}}
\nc{\sU}{\mathsf{U}}\nc{\sV}{\mathsf{V}}\nc{\sW}{\mathsf{W}}\nc{\sX}{\mathsf{X}}
\nc{\sY}{\mathsf{Y}}\nc{\sZ}{\mathsf{Z}}

\newcommand{\mathset}[1]{\left\{#1\right\}}

\newcommand{\abs}[1]{\left|#1\right|}
\newcommand{\ceilenv}[1]{\left\lceil #1 \right\rceil}
\newcommand{\floorenv}[1]{\left\lfloor #1 \right\rfloor}
\newcommand{\parenv}[1]{\left( #1 \right)}
\newcommand{\sparenv}[1]{\left[ #1 \right]}
\newcommand{\bracenv}[1]{\left\{ #1 \right\}}
\nc{\set}[1]{\llbracket #1 \rrbracket}

\newcommand{\Syn}[1]{\textup{Syn}\left(#1\right)}

\newcommand{\dvt}{\mathsf{DVT}}

\newcommand{\tabincell}[2]{\begin{tabular}{@{}#1@{}}#2\end{tabular}}

\newcommand{\rcomment}[1]{\textcolor{red}{#1}}

\title{Codes Correcting Two Bursts of Exactly $b$ Deletions}
\author{Zuo~Ye, Yubo~Sun, Wenjun~Yu, Gennian~Ge and Ohad~Elishco% 
\thanks{This project was supported by the National Key Research and Development Program of China under Grant 2020YFA0712100, the National Natural Science Foundation of China under Grant 12231014 and Grant 12501466, Beijing Scholars Program, and the Israel Science Foundation (Grant No. 1789/23).

Z. Ye is with the Institute of Mathematics and Interdisciplinary Sciences, Xidian University, Xi'an 710126, China. Email: yezuo@xidian.edu.cn.

W. Yu and O. Elishco are with the School of Electrical and Computer Engineering, Ben-Gurion University of the Negev, 
Beer Sheva, Israel. Emails: ohadeli@bgu.ac.il, wenjun@post.bgu.ac.il.

Y. Sun and G. Ge are with the School of Mathematical Sciences, Capital Normal University, Beijing 100048, China. Emails: 2200502135@cnu.edu.cn, gnge@zju.edu.cn. 
}
}

\begin{document}
\maketitle

\begin{abstract}
    In this paper, we investigate codes designed to correct two bursts of deletions, where each burst has a length of exactly $b$, where $b>1$. The previous best construction, achieved through the syndrome compression technique, had a redundancy of at most $7\log n+O\parenv{\log n/\log\log n}$ bits. In contrast, our work introduces a novel approach for constructing $q$-ary codes that attain a redundancy of at most $5\log n+O(\log\log n)$ bits for all $b>1$ and $q\ge2$. Additionally, for the case where $b=1$, we present a new construction of $q$-ary two-deletion correcting codes with a redundancy of $5\log n+O(\log\log n)$ bits, for all $q>2$.
\end{abstract}


\section{Introduction}
%%%%%%%%%%%%%%%%%%%%%%%%%%%%%%%%%%%%%%%%%%%%%%%%%%%%
\IEEEPARstart{A} subset $\cC\subseteq\{0,1,\ldots,q-1\}^n$ (where $q\ge2$) is called a $t$-deletion correcting code, if it has the property that if a codeword $\bfx\in\cC$ is corrupted by deleting $t$ symbols to obtain a subsequence $\bby\in\{0,1,\ldots,q-1\}^{n-t}$, then one can recover $\bfx$ from $\bby$. The study of deletion correcting codes has a long history, dating back to at least the 1960s \cite{Sellers1962}. The seminal work in this field is \cite{VL1966}, whereby proposing a linear-time decoding algorithm, Levenshtein proved that the binary code (Varshamov-Tenengolts code, or VT code for short) constructed in \cite{VT1965} can combat a single deletion error. In 1984, by leveraging the VT code, Tenengolts constructed a non-binary code (Tenengolts code) that can correct a single deletion \cite{Tenengolts1984}. For fixed $t$, $q$ and growing $n$, which is the regime of interest in this paper, the optimal redundancy of a $t$-deletion correcting code $\cC$ of length $n$, defined as $\log\parenv{q^n/\abs{\cC}}$\footnote{All logarithms in this paper are to the base $2$.}, is asymptotically lower bounded by $t\log n+o(\log n)$ \cite{Levenshtein2002ISIT} (for $q=2$, the lower bound is $t\log n+\Omega(1)$ \cite{VL1966}) and upper bounded by $2t\log n-\log\log n+O(1)$ \cite{Alon2024IT}. This implies that the VT code in \cite{VT1965} and the Tenengolts code \cite{Tenengolts1984} have redundancy optimal up to a constant.

Due to applications in DNA-based data storage \cite{Yazdi2015IEEE,hec2019,Dong2020}, document exchange \cite{Cheng2018FOCS,Haeupler2019FOCS}, multiple-deletion correcting codes with low redundancy have attracted a lot of interest in recent years \cite{Cheng2018FOCS,Joshua2018it,Haeupler2019FOCS,Sima2019isit,Gabrys2019it,Sima2020isit_deletion,Bruck2020isit,Sima2020it,Sima2021it,Guruswami2021it,Song2022IT,Song2023IT,ShuLiu2024IT,Tuan2024IT}. To the best of our knowledge, the best known binary $2$-deletion correcting code with redundancy $4\log n+O(\log\log n)$ was given in \cite{Guruswami2021it,Sun2024IT}. For general $t\ge 3$, the smallest redundancy, which is $(4t-1)\log n+o(\log n)$, was achieved by a construction given in \cite{Song2022IT}. For non-binary alphabets, Sima \textit{et al} \cite{Sima2020isit_deletion} presented a family of $q$-ary $t$-deletion correcting codes with redundancy $4t\log n+o(\log n)$ by using the syndrome compression technique. The syndrome compression technique was improved in \cite{Song2022IT} to the so-called syndrome compression technique with pre-coding. A straightforward application of this method will give a $q$-ary $t$-deletion correcting code with redundancy $(4t-1)\log n+o(\log n)$, which is the best-known result in redundancy. When $q>2$ is even and $t=2$, Song and Cai recently constructed a class of $q$-ary $2$-deletion correcting codes with redundancy $5\log n+O(\log\log n)$ \cite{Song2023IT}. And in a following work \cite{ShuLiu2024IT} the authors presented a $q$-ary $2$-deletion correcting code with redundancy $5\log n+O(\log\log n)$ for all $q>2$. When $t=1$, Nguyen \textit{et al} recently constructed a new $q$-ary single-deletion correcting code with redundancy $\log n+\log q$ \cite{Tuan2024IT}. In addition, they showed that there is a linear time encoder with near-optimal redundancy for their code.

If deletions occur at consecutive positions, we call them a burst of deletions. Codes correcting this type of error are of interest due to applications in DNA-based data storage \cite{Yazdi2015IEEE,Christopher2013}, wireless
sensor networks, and satellite communication devices \cite{Jeong2003}. A code is called a $b$-burst-deletion correcting code, if it can correct any single burst of \emph{exactly} $b$ deletions. In 1970, Levenshtein presented a class of binary codes with redundancy at most $\log n+1$ when $b=2$ \cite{levenvstein1970burst}. For $b\ge3$, Cheng \textit{et al} in 2014 constructed a class of binary codes with redundancy $b\log(n/b+1)$ \cite{LingChengisit}, which was later improved to $\log n+(b-1)\log\log n+O(1)$ by Schoeny \textit{et al} in 2017 \cite{Schoeny2017it}. Schoeny's result was generalized to non-binary alphabets in \cite{Schoeny2017,Saeki2018ISITA}. The best known redundancy for all $q\ge2$ is $\log n+O(1)$, which was contributed recently by Sun \textit{et al}\cite{Yubo2025IT}. It was proved in \cite{Schoeny2017it,Saeki2018ISITA} that the redundancy of a $b$-burst-deletion correcting code is at least $\log n+\Omega(1)$. Therefore, the codes in \cite{levenvstein1970burst} and \cite[Theorem 9]{Yubo2025IT} have redundancy optimal up to a constant.
There are also a lot of works on codes correcting single burst of \emph{at most} $b$ deletions \cite{levenvstein1970burst,Schoeny2017it,Ryan2018IT,JinRyan2020isit,Lenz2020ISIT,Song2023IT,Shuche2024IT,Song2025Entropy,Yubo2025IT}. For readers' convenience, we summarize previous results on codes correcting bursts of deletions in \Cref{tab_cases}.

In this work, we focus on codes correcting two bursts of deletions, where each burst is of length exactly $b$. We call such codes $2$-$b$-burst correcting codes. To the best of our knowledge, there are no explicit results about such codes. A related result can be found in a work of Sima \textit{et al} \cite[Section IV-B]{JinRyan2020isit}, where they considered a more generalized type of burst error pattern: $t$ bursts each of length at most $t_L$ where the deletions in each burst need not occur consecutively ($(t,t_L)$ burst deletions, for short). Let $t=2$ and $t_L=b$. Then their result gives a binary $2$-$b$-burst-deletion correcting code with redundancy at most $8\log n+o(\log n)$. A straightforward application of the syndrome compression technique with pre-coding incurs a code with redundancy at most $7\log n+O(\log n/\log\log n)$, for all $q\ge 2$. This conclusion also holds when $b=1$, i.e., for the case of two-deletion correcting codes. On the other hand, \cite{Sima2020it,Guruswami2021it,Song2023IT,ShuLiu2024IT,Sun2024IT} already confirmed that there are two-deletion correcting codes outperforming the one given by the syndrome compression technique. Specifically, \cite{Sima2020it} and \cite{Guruswami2021it,Sun2024IT} presented binary codes with redundancy at most $7\log n+O(1)$ and $4\log n+O(\log\log n)$, respectively. For non-binary codes, the best-known redundancy is $5\log n+O(\log\log n)$ \cite{Song2023IT,ShuLiu2024IT}. Note that in a burst of deletions, all deletions occur consecutively. Therefore, it is reasonable to deem by intuition that there is no big difference between two-deletion correcting codes and $2$-$b$-burst-deletion correcting codes (where $b>1$). This raises a natural question: for $b>1$, is there a construction that leads to codes that are as good as, or even better than, the one given by the syndrome compression technique? Motivated by this question, in this paper, we investigate new constructions of codes for correcting two $b$-burst-deletions for all $b\ge2$.
Our contributions include:
\begin{itemize}
    \item We establish lower and upper bounds on the size (or equivalently, the redundancy) of $2$-$b$-burst-deletion correcting codes;
    \item A binary $2$-$b$-burst-deletion correcting code of length $n$ with redundancy at most $5\log n+14b\log\log n+O(1)$, for any $b>1$;
    \item A $q$-ary $2$-$b$-burst-deletion correcting code of length $n$ with redundancy at most $5\log n+(14b\ceilenv{\log q}+14)\log\log n+O(1)$, for any $q>2$ and $b>1$;
    \item A new construction of $q$-ary two-deletion correcting codes of length $n$ with redundancy at most $5\log n+(14\ceilenv{\log q}+11)\log\log n+O(1)$, for any $q>2$.
\end{itemize}
Here, it is assumed that $q$ and $b$ are constants with respect to $n$. Therefore, our results show that for $2$-$b$-burst-deletion correcting codes, we can do almost as well as two-deletion correcting codes.

\begin{table*}[!t]
	\centering
	\caption{Previous codes correcting bursts of deletions and corresponding methods}
	\label{tab_cases}
	\begin{tabular}{c|c|c|l}
		\hline
		\hline
		&Redundancies of $q$-ary Codes& References&\quad\quad\quad Core Methods\\
		\hline
		\multirow{10}{*}{\tabincell{c}{single burst\\of size\\exactly $b$}}&\tabincell{c}{$b\log(n/b+1)$\\($q=2$)}&\cite{LingChengisit}&\tabincell{l}{1. representing each codeword as an array with $b$ rows\\2. imposing a VT constraint on each row}\\
		\cline{2-4}
        &\tabincell{c}{$\log n+(b-1)\log\log n+O(1)$\\($q=2$, $q>2$)}&\tabincell{c}{\cite{Schoeny2017it}\\\cite{Schoeny2017}}&\tabincell{l}{1. representing each codeword as an array with $b$ rows\\2. encoding the first row with a VT code with run-length limited constraint\\3. encoding the rest rows with shifted VT codes}\\
		\cline{2-4}
        &$\log n+O(1)$ ($q\ge2$)&\cite{Yubo2025IT}&\tabincell{l}{1. representing each codeword as an array with $b$ rows\\2. representing this array as a $q^b$-ary vector\\3. imposing two types of sum constraints on this vector\\4. consider the signature of this vector\\5. imposing a VT-type constraint together with three types of sum constraints\\\quad on the signature}\\
        \hline
       \multirow{37}{*}{\tabincell{c}{single burst\\of size\\at most $b$}} &$\log n+1$ ($b=2,q=2$)&\tabincell{c}{\cite{levenvstein1970burst}\\\cite{Shuche2024IT}}&\tabincell{l}{1. imposing a sum constraint on rank sequences of codewords\\ \quad\quad\quad\quad\quad or\\2. imposing a VT-type constraint on differential sequences of codewords}\\
        \cline{2-4}
        &\tabincell{c}{$(b-1)\log n+\parenv{\binom{b}{2}-1}\log\log n$\\\quad\quad\quad\quad\quad\quad\quad\quad\quad\quad\quad\quad$+O(1)$\\($q=2$)}&\cite{Schoeny2017it}&\tabincell{l}{\quad extension of the construction of $b$-burst-deletion correcting codes\\\quad in the same work}\\
        \cline{2-4}
        &\tabincell{l}{$\ceilenv{\log b}\parenv{\log n+\binom{b+1}{2}\log\log n}$\\\quad\quad\quad\quad\quad\quad\quad\quad\quad$+O(1)$\\($q=2$)}&\cite{Ryan2018IT}&\tabincell{l}{1. denoting each $t\in[1,b]$ as $t=2^i\cdot j$, where $i\ge0$ and $j$ is odd\\2. representing each code word as an array with $2^i$ rows\\3. imposing a VT-type constraint alongside a ``balanced" constraint on the\\\quad first row to approximate the error positions\\4. representing each codeword as an array with $t$ rows and encoding\\\quad each row with a shifted VT code}\\
        \cline{2-4}
        &\tabincell{c}{$\log n+\binom{b+1}{2}\log\log n+O(1)$\\($q=2$)}&\cite{Lenz2020ISIT}&\tabincell{l}{1. each codeword $\bfx$ is required to be $(\bm{p},\delta)$-dense\\2. associating with $\bfx$ a vector of integers $a_{\bm{p}}(\bfx)$\\3. imposing a constraint on the number of $\bm{p}$ in $\bfx$ and a VT-type constraint\\
        \quad on $a_{\bm{p}}(\bfx)$ to approximate positions of errors\\5. for each $1\le t\le b$, representing $\bfx$ as an array with $t$ rows and\\
        \quad encoding each row with shifted VT codes}\\
        \cline{2-4}
        &$4\log n+o\parenv{\log n}$&\cite{JinRyan2020isit}&syndrome compression technique\\
        \cline{2-4}
        &\tabincell{c}{$\log n+8\log\log n+o\parenv{\log\log n}$\\($q=2$)}&\multirow{5}{*}{\cite{Song2023IT}}&\tabincell{l}{1. applying the same method in \cite{Lenz2020ISIT} to approximate positions of errors\\2. applying a code with $4\log n+o(\log n)$ redundant bits to correct burst\\\quad deletions in short intervals}\\
        \cline{2-2}\cline{4-4}
        &\tabincell{c}{$\log n+(8\log q+8)\log\log n$\\\quad\quad\quad\quad\quad\quad$+o\parenv{\log q\log\log n}$\\($q>2$ is even)}&&\tabincell{l}{1. representing each codeword as a binary array with $\ceilenv{\log q}$ rows\\2. encoding the first row with the binary code in the same work\\3. applying a code with $4\log n+o(\log n)$ redundant bits to correct\\\quad errors in remaining rows}\\
        \cline{2-4}
        &\tabincell{c}{$\log n+8\log\log n+o\parenv{\log\log n}$\\($q\ge2$)}&\cite{Song2025Entropy}&\quad generalization of binary codes in \cite{Song2023IT}\\
        \cline{2-4}
        &\tabincell{c}{$\log n+\log q\log\log n+O(1)$\\($b=2$, $q>2$ is even)}&\cite{Shuche2024IT}&\tabincell{l}{1. representing each codeword as a binary array with $\ceilenv{\log q}$ rows\\2. encoding the first row with the binary code in the same work\\\quad with additional pattern-limited constraint\\3. applying a $P$-bounded version of the binary code to remaining rows}\\
        \cline{2-4}
        &\tabincell{c}{$\log n+b\log\log n+O(1)$\\($q\ge2$)}&\cite{Yubo2025IT}&\tabincell{l}{1. associating each codeword with a binary sequence\\2. applying the same method in \cite{Lenz2020ISIT} to the binary sequence to approximate\\\quad error positions\\3. for each $1\le b^\prime\le b$, applying a bounded $b^\prime$-burst-deletion correcting code\\\quad to correct errors in short intervals}\\
        \hline
        \tabincell{c}{$(t,t_L)$\\burst\\deletions}&\tabincell{c}{$4t\log n+o(\log\log n)$\\($q=2$)}&\cite{JinRyan2020isit}&\tabincell{l}{1. $t$-mixed sequences\\2. syndrome compression technique}\\
        \hline
        
        \hline
        \hline
	\end{tabular}
\end{table*}

The rest of this paper is organized as follows. 
In \Cref{sec_pre}, we introduce some necessary definitions and related results. In 
\Cref{sec_bounds}, we bound above and below the size (or equivalently, the redundancy) of codes correcting two bursts of exactly $b$ deletions. \Cref{sec_qburst} deals with codes for correcting two $b$-burst-deletions. In \Cref{sec_qtwodeletion}, we give a new construction of non-binary two-deletion correcting codes. Finally \Cref{sec_conclusion} concludes this paper.

\section{Preliminaries}\label{sec_pre}
%%%%%%%%%%%%%%%%%%%%%%%%%%%%%%%%%%%%%
In this section, we introduce some necessary definitions, auxiliary conclusions and related results.

For an integer $q\ge2$ and a positive integer $n$, denote $\Sigma_q=\mathset{0,1,\ldots,q-1}$ and $\Sigma_q^n$ the set of all $q$-ary sequences with $n$ symbols. Let $\bfx\in\Sigma_q^n$ be a sequence. Unless otherwise stated, the $i$th coordinate of $\bfx$ is denoted by $x_i$, i.e., $\bfx=x_1\cdots x_n$. We call $n$ the length of $\bfx$ and denote $\abs{\bfx}=n$. For a finite set $A$, we denote by $\abs{A}$ the cardinality of $A$.

For two integers $m$ and $n$ such that $m\le n$, let $[m,n]$ denote the set $\mathset{m,m+1,\ldots,n}$. If $m=1$, denote $[n]=[1,n]$. For a sequence $\bfx\in\Sigma_q^n$ and a subset $I=\mathset{i_1,i_2,\ldots,i_t}\subseteq[n]$ where $i_1<i_2<\cdots<i_t$, we define $\bfx_{I}\triangleq x_{i_1}x_{i_2}\cdots x_{i_t}$. For each $I\subseteq[n]$, we say $\bfx_{I}$ is a \textbf{subsequence} of $\bfx$. In particular, if $I$ is an \textbf{interval} of $[n]$ (i.e., $I=[i,j]$ for some $1\le i\le j\le n$), we say $\bfx_{I}$ is a \textbf{substring} of $\bfx$. A \textbf{run} in $\bfx$ is a maximal substring consisting of the same symbols. The number of runs of $\bfx$ is denoted by $r(\bfx)$. For example, if $\bfx=100101$, then there are five runs in $\bfx$: $1$, $00$, $1$, $0$ and $1$. So $r\parenv{\bfx}=5$. 

The concatenation of two sequences $\bfx$ and $\bby$ is denoted by $\bfx\bby$. For example, let $\bfx=102$ and $\bby=121$ be two sequences in $\Sigma_3^3$, then $\bfx\bby=102121\in\Sigma_3^6$. 
Let $b$ and $n$ be two positive integers satisfying $b<n$. When a substring of length $b$ is deleted, we refer to it as a \textbf{deletion-burst of size $b$} or a \textbf{$b$-burst-deletion}; that is to say, from $\bfx\in\Sigma_q^n$, we obtain a subsequence $\bfx_{[n]\setminus[i,i+b-1]}$ for some $1\le i\le n-b+1$.

In this paper, we focus on codes correcting two $b$-burst-deletions. Suppose $\bfx\in\Sigma_q^n$, where $n>2b$. There are two ways to define two $b$-burst-deletions:
\begin{enumerate}[(\textbf{D}1)]
    \item the two bursts are caused by two channels: $\bfx$ passes the first channel, resulting in $\bbz=\bfx_{[n]\setminus[i_1,i_1+b-1]}$ and then $\bbz$ passes the second channel, resulting in $\bby=\bbz_{[n-b]\setminus[i_2,i_2+b-1]}$;
    \item the two bursts are caused by a single channel: symbols in $\bfx$ pass a channel sequentially and we receive
    $\bby=\bfx_{[n]\setminus I_1\cup I_2}$, where $I_1$ and $I_2$ are two disjoint intervals of length $b$ in [n].% This definition is motivated by the nanopore sequencing technology, in which a DNA strand pass a nanoscale hole  nucleotide by nucleotide \cite{Omer2024TMBMSC,Yunhao2021NB}. So we can assume that the two bursts do not overlap.
\end{enumerate}

\begin{remark}
There is another possibility: the two bursts might overlap and result in a single burst that is shorter than $2b$. We do not take this situation into account, since it is covered by a more comprehensive problem: correcting two bursts of deletions, where each burst has length \emph{at most} $b$. Our idea in this paper fails in this situation. We left this problem for future research.
\end{remark}

In fact, (D1) and (D2) are equivalent. Firstly, it is clear that (D1) covers (D2). Next, we show that (D2) also covers (D1). 
\begin{observation}\label{obs_disjoint}
    Let $n>2b$. Suppose $\bfx\in\Sigma_q^n$ and $\bby$ is obtained from $\bfx$ by process (D1). Then there exist two intervals $I_1=[j_1,j_1+b-1]$, $I_2=[j_2,j_2+b-1]\subseteq[n]$, where $j_2-j_1\ge b$, such that $\bby=\bfx_{[n]\setminus(I_1\cup I_2)}$. In particular, if $\bby$ is obtained from $\bfx$ by two $b$-burst-deletions, we can always assume that $\bby$ is obtained from $\bfx$ by deleting two non-overlapping substrings of length $b$ from $\bfx$.
\end{observation}
\begin{IEEEproof}
    By assumption, there is $1\le i_1\le n-b+1$ and $1\le i_2\le n-2b+1$ such that $\bby=\bbz_{[n-b]\setminus[i_2,i_2+b-1]}$, where $\bbz=\bfx_{[n]\setminus[i_1,i_1+b-1]}$. If $i_2\ge i_1$, let $j_1=i_1$ and $j_2=i_2+b$. Then the conclusion follows.

    Now suppose $1\le i_2\le i_1-1$. If $i_2\le i_1-b$, let $j_1=i_2$ and $j_2=i_1$. Then the conclusion follows. If $i_1-b<i_2<i_1$, it is clear that $\bby=\bfx_{[n]\setminus[i_2,i_2+2b-1]}$. Let $j_1=i_2$ and $j_2=i_2+b$. Then the conclusion follows.
\end{IEEEproof}

For $t\in\{1,2\}$ and $n>tb$, define
$$
\cB_t^b\parenv{\bfx}=\mathset{\bby\in\Sigma_q^{n-tb}:
\begin{array}{c}
     \bby\text{ is obtained from }\bfx\\
     \text{by }t\; b\text{-burst-deletion(s)}
\end{array}
}.
$$
When $b=1$, we use notation $\cB_t(\bfx)$ instead of $\cB_t^b\parenv{\bfx}$.
\begin{definition}
  Let $\cC$ be a subset of $\Sigma_q^n$ with $\abs{\cC}\ge2$. Suppose $t\in\{1,2\}$. We call $\cC$ a \textbf{$t$-$b$-burst-deletion correcting code} if 
  $\cB_t^b(\bfx)\cap\cB_t^b(\bby)=\emptyset$ for any two distinct $\bfx,\bby\in\cC$. In particular, if $b=1$, we call $\cC$ a $t$-deletion correcting code.
\end{definition}

Clearly, if any $\bfx\in\cC$ can be uniquely and efficiently recovered from any given $\bfx^\prime\in\cB_t^b(\bfx)$, then $\cC$ is a $t$-$b$-burst-deletion correcting code. Here, ``efficiently" means that the time complexity of decoding $\bfx$ from $\bby$ is polynomial in $n$. In this paper, we construct $2$-$b$-burst-deletion correcting codes and show that any codeword can be uniquely and efficiently decoded.

The \textbf{redundancy} of a code $\cC\subseteq\Sigma_q^n$ is defined to be $\rho(\cC)=\log\parenv{q^n/\abs{\cC}}$. All logarithms in this paper are to the base $2$. In addition, we always assume that $q$ and $b$ are fixed with respect to the code-length $n$.

Let $n^{\prime}$ and $n$ be two positive integers satisfying $n^{\prime}< n$. 
For each sequence $\bfx\in\Sigma_q^n$, let $\widetilde{\bfx}$ be the zero padding of $\bfx$ to the shortest length that is greater than $n$ and is divisible by $n^{\prime}$, that is, $\widetilde{\bfx}=\bfx0^{\ceilenv{n/n^\prime}n^\prime-n}$, and then $\abs{\widetilde{\bfx}}$ is divided by $n^\prime$. 
We can represent $\bfx$ as an $n^{\prime}\times\ceilenv{n/n^\prime}$ array $A\parenv{\bfx,n^{\prime}}=\sparenv{a_{i,j}}$, where $a_{i,j}=\tilde{x}_{i+n^{\prime}j}$ for all $1\le i\le n^{\prime}$ and $0\le j\le\ceilenv{n/n^\prime}-1$. In other words, the $i$-th row of $A\parenv{\bfx,n^{\prime}}$ is
$$
A\parenv{\bfx,n^{\prime}}_i\triangleq\parenv{\tilde{x}_i,\tilde{x}_{i+n^{\prime}},\tilde{x}_{i+2n^{\prime}},\ldots,\tilde{x}_{i+n^{\prime}\parenv{\ceilenv{\frac{n}{n^{\prime}}}-1}}}.
$$
We call $A\parenv{\bfx,n^{\prime}}$ a matrix (or array) representation of $\bfx$.
If $n^{\prime}$ is clear from the context, we will denote $A\parenv{\bfx,n^{\prime}}$ by $A(\bfx)$.

For example, let $n=7$, $\bfx=1011010\in\Sigma_2^{7}$ and $n^{\prime}=2$. 
Then $\widetilde{\bfx}=10110100\in\Sigma_2^{8}$ and 
$$
A\parenv{\bfx,2}=
\begin{pmatrix}
    1&1&0&0\\
    0&1&1&0
\end{pmatrix}.
$$
If $n^\prime=3$, then $\widetilde{\bfx}=\bfx00$ and so
$$
A\parenv{\bfx,3}=
\begin{pmatrix}
    1&1&0\\
    0&0&0\\
    1&1&0
\end{pmatrix}.
$$

In this paper, when dealing with $t$-$b$-burst-deletion correcting codes, it is helpful to represent a sequence $\bfx$ of length $n$ as matrix $A(\bfx,b)$. To avoid ceiling functions (for example, $\ceilenv{n/b}$), we always assume $b\mid n$. All results in this paper still hold even if $b\nmid n$, as long as we replace $n/b$ by $\ceilenv{n/b}$. Throughout this paper, for a matrix $A$, denote by $A_i$ and $A_{i,j}$ the $i$-th row of $A$ and the entry in the $i$-th row and $j$-th column, respectively.

The next observation is a straightforward result of \Cref{obs_disjoint}. 
\begin{observation}\label{obs_deletionposition}
    For each $b<n$ and $b\mid n$, we can represent a sequence $\bfx\in\Sigma_q^n$ as a $b\times n/b$ array $A(\bfx)$. Any two $b$-burst-deletions in $\bfx$ will induce two deletions in each row of $A(\bfx)$. Furthermore, if the positions of the two deletions in $A(\bfx)_1$ are $j_1$ and $j_2$, then for each $2\le i\le b$, one of the two deletions in $A(\bfx)_i$ occurred at coordinate $j_1-1$ or $j_1$, and the other deletion occurred at coordinate $j_2-1$ or $j_2$.
\end{observation}

\subsection{Related Results}
Recently, there have been two notable developments in non-binary two-deletion correcting codes with low redundancy. In \cite[Theorem 1]{Song2023IT}, the authors presented a $q$-ary two-deletion correcting code with redundancy at most $5\log n+(16\log q+10)\log\log n+o(\log\log n)$, for any \emph{even} $q>2$. 
In \cite{ShuLiu2024IT}, a $q$-ary two-deletion correcting code was constructed, with redundancy at most $5\log n+10\log\log n+O_q(1)$ (where $O_q(1)$ denotes a constant depending only on $q$), for any $q>2$. In \Cref{sec_qtwodeletion}, we present a new construction of $q$-ary two-deletion correcting codes with redundancy at most $5\log n+(14\ceilenv{\log q}+11)\log\log n+O_q(1)$, for any $q>2$.

Regarding codes that can correct two $b$-burst-deletions (where $b>1$), there are, to our knowledge, no explicit results available. A related result can be found in the work of Sima \textit{et al} \cite[Section IV-B]{JinRyan2020isit}, where they considered a more generalized type of burst error pattern: $t$ bursts each of length at most $t_L$ with the deletions in each burst not necessarily occurring consecutively. For $t=2$ and $t_L=b$, their result provides a binary $2$-$b$-burst-deletion correcting code with redundancy at most $8\log n+o(\log n)$. Their result was derived using the syndrome compression technique, which was later extended to syndrome compression with pre-coding in \cite{Song2022IT}. We will apply this extended technique to provide a $q$-ary code with redundancy at most $7\log n+O\parenv{\log n/\log\log n}$, for all $q\ge 2$.

For a subset $\cE\subseteq\Sigma_q^n$ and a sequence $\bfx\in\cE$, define
$$
\cN_{\cE}\parenv{\bfx}=\mathset{\bby\in\cE~:~\bby\ne\bfx\text{ and }\cB_2^b\parenv{\bby}\cap\cB_2^b\parenv{\bfx}\ne\emptyset}.
$$
In other words, $\cN_{\cE}\parenv{\bfx}$ is the set of all sequences (except $\bfx$) in $\cE$ whose error-ball intersects with that of $\bfx$.
\begin{lemma}\label{lem_sycompress}\cite{Sima2019isit,Song2022IT,Zuo2024IT}
    Let $\cE\subseteq\Sigma_q^n$ be a code and $N>\max\bracenv{\abs{\cN_{\cE}\parenv{\bfx}}~:~\bfx\in\cE}$.
Suppose that the function $f:\Sigma_q^n\rightarrow\{0,1\}^{R(n)}$ (where $R(n)$ is a function of $n$ and $R(n)\ge2$) satisfies the following property:
\begin{enumerate}[$(\textup{\textbf{P}} 1)$]
    \item if $\bfx\in\Sigma_q^n$ and $\bby\in\cN_{\Sigma_q^n}\parenv{\bfx}$, then $f\parenv{\bfx}\ne f\parenv{\bby}$.
\end{enumerate}
Then there exists a function $\bar{f}:\cE\rightarrow\{0,1\}^{2\log(N)+O\parenv{\frac{R(n)}{\log\parenv{R(n)}}}}$, computable in polynomial time\footnote{In this paper, when saying that a function is computable in polynomial/linear time, we mean that this function is computable in time polynomial/linear in the code-length $n$.} such that $\bar{f}\parenv{\bfx}\ne \bar{f}\parenv{\bby}$ for any $\bfx\in\cE$ and $\bby\in\cN_{\cE}\parenv{\bfx}$.
\end{lemma}

Let $\cE$ be a $1$-$b$-burst correcting code and $\bar{f}$ be given in \Cref{lem_sycompress}. Then \Cref{lem_sycompress} asserts that if $\bfx,\bby\in\cE$ are distinct codewords and $\bar{f}(\bfx)=\bar{f}(\bby)$, we have $\cB_2^b(\bfx)\cap\cB_2^b(\bby)=\emptyset$. Therefore,
for any $\bba\in\{0,1\}^{2\log(N)+O\parenv{\frac{R(n)}{\log\parenv{R(n)}}}}$, 
the code $\cE^\prime=\mathset{\bfx\in\cE:\bar{f}(\bfx)=a}$ is a $2$-$b$-burst-deletion correcting code. 
Furthermore, by the pigeonhole principle, there exists an $\bba$ such that the redundancy of $\cE^\prime$ is at most $\rho\parenv{\cE^\prime}=\rho(\cE)+2\log(N)+O\parenv{\frac{R(n)}{\log\parenv{R(n)}}}$.\footnote{We select $\cE$ to be a $1$-$b$-burst correcting code to obtain a better redundancy. If we take $\cE=\Sigma_q^n$ we get a redundancy of $8\log n+o(\log n)$.}

For the choice of $\cE$, we have the following result.
\begin{lemma}\cite[Theorem 9, $t=b,s=0$]{Yubo2025IT}\label{lem_singleburst}
    For all $q\ge 2$ and $n\ge b$, there is a function $\phi:\Sigma_q^n\rightarrow\Sigma_2^{\log n+O_{q,b}(1)}$, computable in linear time, such that for any $\bfx\in\Sigma_q^n$, given $\phi(\bfx)$ and $\bby\in\cB_1^b\parenv{\bfx}$, one can uniquely and efficiently recover $\bfx$. Here, $O_{q,b}(1)$ is a constant dependent only on $q$ and $b$.
\end{lemma}

This lemma gives a $1$-$b$-burst-deletion correcting code $\cE$ with redundancy $\log n+O_{q,b}(1)$. Since $\cE$ can correct single $b$-burst-deletion, by a simple counting, we can see that $\abs{\cN_{\cE}(\bfx)}< q^bn^3$. In fact, each codeword in $\cN_{\cE}(\bfx)$ can be obtained in the following three steps:
\begin{enumerate}[1)]
    \item Delete two substrings of length $b$ from $\bfx$, resulting in $\bbz^{(1)}$. There are less than $n^2$ possibilities for $\bbz^{(1)}$.
    \item For each $\bbz^{(1)}$, insert a sequence of length $b$ into $\bbz^{(1)}$ and get a sequence $\bbz^{(2)}\in\Sigma_q^{n-b}$. For each $\bbz^{(1)}$, there are at most $q^bn$ possibilities for $\bbz^{(2)}$.
    \item Insert a sequence of length $b$ into $\bbz^{(2)}$ to get a sequence
    $\bby\in\Sigma_q^n$. Since $\cE$ is a $1$-$b$-burst-deletion correcting code, for each $\bbz^{(2)}$, there is at most one $\bby$ which is in $\cN_{\cE}(\bfx)$.
\end{enumerate}

Therefore, we can let $N=q^bn^3$. Now the redundancy of $\cE^\prime$ is at most $7\log n+O\parenv{\frac{R(n)}{\log\parenv{R(n)}}}$. To conclude our discussion, it remains to find an $f$ satisfying (P1) in \Cref{lem_sycompress} such that $R(n)=O(\log n)$, which is given in \Cref{lem_trivial2burst}. The proof of \Cref{lem_trivial2burst} is based on \Cref{lem_twodeletion,lem_trivial2deletion}.

The following lemma is a corollary of \cite[Theorem 2]{Sima2020it}.
\begin{lemma}\cite[Theorem 2]{Sima2020it}\label{lem_twodeletion}
    For any integer $n\ge 3$, there exists a function $\xi:\Sigma_2^n\rightarrow\Sigma_2^{7\log n+O\parenv{1}}$, computable in linear time, such that for any $\bfx\in\Sigma_2^n$, given $\xi\parenv{\bfx}$ and any $\bby\in\cB_2\parenv{\bfx}$ (i.e., $\bby$ is obtained from $\bfx$ by two deletions), one can uniquely and efficiently recover $\bfx$.
\end{lemma}

This result can be extended to arbitrary finite alphabets in the following way.
\begin{lemma}\label{lem_trivial2deletion}
    Suppose $q\ge2$. There is a function $\xi_1:\Sigma_q^n\rightarrow\Sigma_2^{7\ceilenv{\log q}\log n+O_q(1)}$, such that for any $\bfx\in\Sigma_q^n$, given $\bby\in\cB_2\parenv{\bfx}$ and $\xi_1(\bfx)$, one can uniquely and efficiently recover $\bfx$. Here, $O_{q}(1)$ is a constant dependent only on $q$.
\end{lemma}
\begin{IEEEproof}
    Any $\bfx\in\Sigma_q^n$ can be uniquely represented as an array
    $$
    M(\bfx)\triangleq
    \begin{pmatrix}
      x_{1,1}&\cdots&x_{1,n}\\
      \vdots&\cdots&\vdots\\
      x_{\ceilenv{\log q},1}&\cdots&x_{\ceilenv{\log q},n}
    \end{pmatrix},
    $$
    where $x_{k,i}\in\{0,1\}$ such that $x_i=\sum_{k=1}^{\ceilenv{\log q}}x_{k,i}2^{k-1}$ for all $1\le i\le n$.

Denote the $k$-th row of $M(\bfx)$ by $M(\bfx)_k$. Suppose $\bby\in\cB_2(\bfx)$. It is clear that $M(\bby)_k\in\cB_2\parenv{M(\bfx)_k}$ for all $1\le k\le\ceilenv{\log q}$. Let $\xi(\cdot)$ be the function defined in \Cref{lem_twodeletion}. For $\bfx\in\Sigma_q^n$, define $\xi_1(\bfx)\triangleq\parenv{\xi\parenv{M(\bfx)_1},\ldots,\xi\parenv{M(\bfx)_{\ceilenv{\log q}}}}$. Then by \Cref{lem_twodeletion},  given $\bby\in\cB_2\parenv{\bfx}$ and $\xi_1(\bfx)$, one can uniquely and efficiently recover $\bfx$. Since each $\xi\parenv{M(\bfx)_k}$ is a binary vector of length $7\log n+O(1)$, we can see that $\xi_1(\bfx)$ is a binary vector of length $7\ceilenv{\log q}\log n+O_q(1)$.
\end{IEEEproof}

\begin{lemma}\label{lem_trivial2burst}
   Suppose $b>1$, $n>2b$ and $q\ge2$. There is a function $\psi:\Sigma_q^n\rightarrow\Sigma_2^{7b\ceilenv{\log q}\log(n/b)+O_{q,b}(1)}$, such that for any $\bfx\in\Sigma_q^n$, given $\bby\in\cB_2^b\parenv{\bfx}$ and $\psi(\bfx)$, one can uniquely and efficiently recover $\bfx$.
\end{lemma}
\begin{IEEEproof}
    Let $A(\bfx)=A(\bfx,b)$ and $A(\bby)=A(\bby,b)$. Since $\bby\in\cB_2^b(\bfx)$, we have $A(\bby)_i\in\cB_2\parenv{A(\bfx)_i}$ for all $1\le i\le b$. Let $\xi_1(\cdot)$ be the function defined in \Cref{lem_trivial2deletion}. Define $\psi(\bfx)\triangleq\parenv{\xi_1\parenv{A(\bfx)_1},\ldots,\xi_1\parenv{A(\bfx)_b}}$. Then by \Cref{lem_trivial2deletion},  given $\bby\in\cB_2^b\parenv{\bfx}$ and $\psi(\bfx)$, one can uniquely and efficiently recover $\bfx$. Since each $\xi_1(A(\bfx)_i)$ is a binary vector of length $7\ceilenv{\log q}\log(n/b)+O_q(1)$, we can see that $\psi(\bfx)$ is a binary vector of length $7b\ceilenv{\log q}\log(n/b)+O_{q,b}(1)$.
\end{IEEEproof}

Taking $f=\psi$ gives a function satisfying (P1) as needed. 

\bigskip
Before proceeding to subsequent sections, we introduce a useful lemma, which will be used in \Cref{sec_qruns}.

Let $N\ge2$ be an integer. Suppose $\bfx\in[0,N-1]^n$, where $n>4$. Let $?$ denote an unknown symbol (not in $[0,N-1]$). If $\bby\in\parenv{[0,N-1]\cup\{?\}}^{n}$ such that $y_i=y_{i+1}=y_{j}=y_{j+1}=?$ and $y_k=x_k$ for any $k\notin\{i,i+1,j,j+1\}$, we say that $\bby$ is obtained from $\bfx$ by two bursts of erasures (of length two). For our purpose in \Cref{sec_qruns}, assume $j\ge i+2$. For a sequence $\bfx$ over the alphabet $[0,N-1]$, denote $\Syn{\bfx}\triangleq\sum_{i=1}^nix_i$. In the following lemma, denote $A(\bfx)=A(\bfx,2)$, i.e., the matrix representation of $\bfx$ with two rows.
\begin{lemma}\label{lem_tberasures}
  For any $0\le a_1,a_2<2N$ and $0\le b<nN^2$, define $\cC$ to be the set of all sequences $\bfx\in[0,N-1]^n$ that satisfies:
    \begin{itemize}
        \item[(C1)] $\sum_{j=1}^{\ceilenv{n/2}}A(\bfx)_{1,j}\equiv a_1\pmod{2N}$, $\sum_{j=1}^{\ceilenv{n/2}}A(\bfx)_{2,j}\equiv a_2\pmod{2N}$;
        \item[(C2)] $W(\bfx)\equiv b\pmod{nN^2}$, where $W(\bfx)=\Syn{A(\bfx)_1}+(2N-1)\cdot\Syn{A(\bfx)_2}$.
    \end{itemize}
    Then the code $\cC$ can correct two bursts (of length two) of erasures. In particular, there is a function $\varphi:[0,N-1]^n\rightarrow\Sigma_2^{\log n+4\log N+2}$, efficiently computable, such that for any $\bfx\in[0,N-1]^n$, given $\varphi(\bfx)$, we can efficiently and uniquely recover $\bfx$ from $\bby$, where $\bby$ is any given sequence obtained from $\bfx$ by two bursts of erasures.
\end{lemma}
\begin{IEEEproof}
Firstly, if the correctness of the code $\cC$ is proved, we can define 
\begin{align*}
\varphi(\bfx)\triangleq&\left(\sum_{j=1}^{\ceilenv{n/2}}A(\bfx)_{1,j}\pmod{2N},\right.\\
&\quad\sum_{j=1}^{\ceilenv{n/2}}A(\bfx)_{2,j}\pmod{2N},\\
&\quad W(\bfx)\pmod{nN^2}\Bigg).
\end{align*}
Here, we view $\varphi(\bfx)$ as a binary vector. There are at most $(2N)^2\cdot(nN^2)$ values of $\varphi(\bfx)$. As a result, the length of $\varphi(\bfx)$, when viewed as a binary vector, is at most $\log n+4\log N+2$. It remains to prove the correctness of $\cC$.

Suppose $\bby$ is obtained from a codeword $\bfx\in\cC$ by two bursts of erasures. Then $A(\bby)_i$ is obtained from $A(\bfx)_i$ by two erasures for each $i=1,2$. Denote the error positions in $A(\bby)_1$ are $i_1,i_2$, and the error positions in $A(\bby)_2$ are $i_3,i_4$, where $i_1< i_2$ and $i_3\le i_4$. Note that $i_1,i_2,i_3$ and $i_4$ are known to us. Clearly, we have $i_3\in\{i_1,i_1-1\}$ and $i_4\in\{i_2,i_2-1\}$. This implies that $i_4-i_3\in\{i_2-i_1,i_2-i_1-1,i_2-i_1+1\}$.

Next, we describe how to decode $\bfx$ from $\bby$. Clearly, it is sufficient to recover the values of $A(\bfx)_{1,i_1}$, $A(\bfx)_{1,i_2}$, $A(\bfx)_{2,i_3}$ and $A(\bfx)_{2,i_4}$. Let
\begin{equation}\label{eq_tberasure1}
\begin{aligned}
    \delta_1=\parenv{a_1-\sum_{j=1,j\ne i_1,i_2}^{\ceilenv{n/2}}A(\bfx)_{1,j}}\pmod{2N},\\
    \delta_2=\parenv{a_2-\sum_{j=1,j\ne i_3,i_4}^{\ceilenv{n/2}}A(\bfx)_{2,j}}\pmod{2N}.
\end{aligned}
\end{equation}
Since $0\le A(\bfx)_{1,i_1}+A(\bfx)_{1,i_2},A(\bfx)_{2,i_3}+A(\bfx)_{2,i_4}<2N$ and $0\le\delta_1,\delta_2<2N$, it follows from Condition (C1) and \Cref{eq_tberasure1} that $\delta_1=A(\bfx)_{1,i_1}+A(\bfx)_{1,i_2}$ and $\delta_2=A(\bfx)_{2,i_3}+A(\bfx)_{2,i_4}$.

For simpler notations, denote $\alpha_1=A(\bfx)_{1,i_1}$ and $\alpha_2=A(\bfx)_{2,i_3}$. Then we have $A(\bfx)_{1,i_2}=\delta_1-\alpha_1$ and $A(\bfx)_{2,i_4}=\delta_2-\alpha_2$. Therefore, it remains to obtain the values of $\alpha_1$ and $\alpha_2$. To that end, let
\begin{equation}\label{eq_tberasure2}
\begin{aligned}
     \Delta=&\left(b-\sum_{j=1,j\ne i_1,i_2}^{\ceilenv{n/2}}j\cdot A(\bby)_{1,j}\right.\\
     &\quad\quad\left.-(2N-1)\sum_{j=1,j\ne i_3,i_4}^{\ceilenv{n/2}}j\cdot A(\bby)_{2,j}\right)\pmod{nN^2}.
\end{aligned}
\end{equation}
\begin{figure*}[!t]
    \begin{equation}\label{eq_tberasure3}
        \begin{aligned}
            \Delta&=\parenv{W(\bfx)-\sum_{j=1,j\ne i_1,i_2}^{\ceilenv{n/2}}j\cdot A(\bfx)_{1,j}-(2N-1)\sum_{j=1,j\ne i_3,i_4}^{\ceilenv{n/2}}j\cdot A(\bfx)_{2,j}}\pmod{nN^2}\\
        &=\parenv{i_1\alpha_1+i_2(\delta_1-\alpha_1)+(2N-1)(i_3\alpha_2+i_4(\delta_2-\alpha_2))}\pmod{nN^2}.
        \end{aligned}
    \end{equation}\hrulefill
\end{figure*}
Since each term in the right-hand side of (\ref{eq_tberasure2}) is known, we can obtain the value of $\Delta$. Furthermore, by (\ref{eq_tberasure2}), Condition (C2), and the relationship between $A(\bfx)$ and $A(\bby)$, we have \eqref{eq_tberasure3}.
%\begin{equation}%\label{eq_tberasure3}
%    \begin{aligned}
%        \Delta&=\parenv{W(\bfx)-\sum_{j=1,j\ne i_1,i_2}^{\ceilenv{n/2}}j\cdot A(\bfx)_{1,j}-(2N-1)\sum_{j=1,j\ne i_3,i_4}^{\ceilenv{n/2}}j\cdot A(\bfx)_{2,j}}\pmod{nN^2}\\
%        &=\parenv{i_1\alpha_1+i_2(\delta_1-\alpha_1)+(2N-1)(i_3\alpha_2+i_4(\delta_2-\alpha_2))}\pmod{nN^2}.
%    \end{aligned}
%\end{equation}

Let $\Delta^\prime=\parenv{i_2\delta_1+(2N-1)i_4\delta_2-\Delta}\pmod{nN^2}$, which can be calculated since values of $\delta_1$, $\delta_2$ and $\Delta$ are known to us. Then it follows from \Cref{eq_tberasure3} that
\begin{equation}\label{eq_tberasure4}
(i_2-i_1)\alpha_1+(2N-1)(i_4-i_3)\alpha_2\equiv\Delta^\prime\pmod{nN^2}.
\end{equation}
Since $i_1<i_2\le\ceilenv{n/2}$ and $i_4-i_3\le\ceilenv{n/2}$, we have $1\le i_2-i_1,i_4-i_3\le n/2$. Combining this with the fact $0\le\alpha_1,\alpha_2\le N-1$, we conclude that $0\le (i_2-i_1)\alpha_1+(2N-1)(i_4-i_3)\alpha_2<nN^2$. Now \Cref{eq_tberasure4} implies
\begin{equation}\label{eq_tberasure5}
(i_2-i_1)\alpha_1+(2N-1)(i_4-i_3)\alpha_2=\Delta^\prime.
\end{equation}

Recall that $i_4-i_3\in\{i_2-i_1-1,i_2-i_1,i_2-i_1+1\}$. By definition, the two bursts of erasures in $\bfx$ do not overlap. It follows that $i_2\ge i_1+2$ if $i_4-i_3=i_2-i_1-1$. Therefore, we always have $i_4-i_3\ge 1$. This implies $(2N-1)(i_4-i_3)>1$. Then it follows from \Cref{eq_tberasure5} that 
\begin{equation}\label{eq_tberasure6}
    (i_2-i_1)\alpha_1\equiv\Delta^\prime\pmod{(2N-1)(i_4-i_3)}.
\end{equation}

Note that $0\le\alpha_1\le N-1$. When $i_4-i_3=i_2-i_1$ or $i_2-i_1+1$, it is easy to see that $0\le(i_2-i_1)\alpha_1<(2N-1)(i_4-i_3)$. When $i_4-i_3=i_2-i_1-1$, since $i_2-i_1\ge 2$, we have $(2N-1)(i_4-i_3)-(i_2-i_1)\alpha_1=(i_2-i_1)(2N-1-\alpha_1)-(2N-1)\ge 2N-1-2\alpha_1>0$ and hence $(i_2-i_1)\alpha_1<(2N-1)(i_4-i_3)$. Now it follows from \Cref{eq_tberasure6} that $(i_2-i_1)\alpha_1=\Delta^{\prime\prime}$, where $\Delta^{\prime\prime}=\Delta^{\prime}\pmod{(2N-1)(i_4-i_3)}$. From this, we get $\alpha_1=\frac{\Delta^{\prime\prime}}{(i_2-i_1)}$. Then by \Cref{eq_tberasure5}, we have $\alpha_2=\frac{\Delta^\prime-(i_2-i_1)\alpha_1}{(2N-1)(i_4-i_3)}$. Now the proof is completed.
\end{IEEEproof}

\section{Bounds}\label{sec_bounds}
%%%%%%%%%%%%%%%%%%%%%%%%%%%%%%%%%%%%%%%%%
We could not find any existing upper or lower bounds on the maximum size of a $2$-$b$-burst-deletion code. In this section, we will derive these bounds.

Let $M_{q,n,b}$ be the maximum size of a $2$-$b$-burst-deletion correcting code in $\Sigma_q^n$, where $n>2b$.
\begin{theorem}\label{thm_lowerbound}
 The maximum size of a $2$-$b$-burst-deletion correcting code satisfies $M_{n,q,b}\ge\frac{q^{n-2b}}{\binom{n}{2}^2}$.
\end{theorem}
\begin{IEEEproof}
    We construct a graph $G$ where the vertex set $V(G)$ is $\Sigma_q^n$ and two \emph{distinct} vertices $\bfx$, $\bby$ in $\Sigma_q^n$ is connected by an edge (denoted by $\bfx\sim\bby$) if and only if $\cB_2^b(\bfx)\cap\cB_2^b(\bby)\ne\emptyset$. An independent set of $G$ is a subset of $\Sigma_q^n$ such that any two distinct vertices are not connected by an edge. Let $\alpha(G)$ be the maximum size of an independent set of $G$. By definition, a subset $\cC\subseteq\Sigma_q^n$ is a $2$-$b$-burst-deletion correcting code if and only if $\cC$ is an independent set in $G$. Therefore, we have $M_{q,n,b}=\alpha(G)$.
    For a vertex $\bfx$, let $d(\bfx)$ be the number of $\bby$ such that $\bfx\sim\bby$.
     Then it follows from \cite[page 100, Theorem  1]{Alon2016} that 
     \begin{equation}\label{eq_lowerbound}
         M_{q,n,b}\ge\sum_{\bfx\in\Sigma_q^n}\frac{1}{d(\bfx)+1}.
     \end{equation}
    
    For $\bfx\in\Sigma_q^n$, by \Cref{obs_disjoint}, we conclude that each $\bby$ (including $\bfx$ itself) with $\cB_2^b(\bfx)\cap\cB_2^b(\bby)\ne\emptyset$ can be obtained as follows: 1) deleting two non-overlap substrings of length $b$ from $\bfx$, resulting in a sequence $\bbz\in\Sigma_q^{n-2b}$; 2) inserting two sequences of length $b$ into $\bbz$. Therefore, we have $d(\bfx)+1\le \binom{n}{2}^2q^{2b}$.
\end{IEEEproof}

In \cite[Section IV-B]{Saeki201804}, the authors proved an upper bound of $1$-$b$-burst-deletion correcting codes. Next, we adapt their idea to derive an upper bound on $M_{q,n,b}$. Recall that $\ln(\cdot)$ is the natural logarithm function.
\begin{theorem}\label{thm_upperbound}
For $q\ge 2$, let $f(q)=\min\mathset{\frac{1}{q},\frac{q-1}{2q},\frac{(q-1)^2}{q^2-3q+6}\parenv{\frac{1}{q}-\frac{(q-1)\ln q}{2q^3}}}$.
If $n\ge30$ is sufficiently large such that $\frac{\log n}{n}<\frac{\log q}{12}f(q)^2$ and $\parenv{1-\frac{q}{q-1}\sqrt{\frac{12\log n}{n\log q}}}^2\parenv{1-b/n}^2\ge 2/3$, the maximum size of a $2$-$b$-burst-deletion correcting code satisfies
$$
M_{q,n,b}\le\parenv{\frac{3b^2}{q^{2b-2}(q-1)^2}+\frac{(1.121)^{3b}}{n}}\frac{q^n}{n^2}.
$$
\end{theorem}
\begin{IEEEproof}
    Define $m = n/b-1$. Let $\cC\subseteq \Sigma_q^n$ be a $2$-$b$-burst-deletion correcting code. Set $\epsilon = \sqrt{\frac{12\log n}{n\log q}}$ and $t =\parenv{1-\frac{1}{q}-\epsilon}m$. We partition  $\cC$ into two disjoint subsets: $\cC = \cC_1 \cup \cC_2$, where $\cC_1 = \mathset{\bfx \in \cC: r(A(\bfx)_i) \ge t+2 \text{ for some } 1\leq i \leq b }$ and $\cC_2 = \mathset{\bfx\in\cC: r(A(\bfx)_i) \le t+1 \text{ for all } 1\le i \le b }$ (recall that $r(\cdot)$ denotes the number of runs). To derive an upper bound of $\abs{\cC}$, it is sufficient to upper bound $\abs{\cC_1}$ and $\abs{\cC_2}$. 
	
    For any $\bfx \in \mathcal{C}$, define $\cA_2^b(\bfx)=\mathset{A(\bfx^\prime):\bfx^\prime\in\cB_2^b(\bfx)}$. Let $A(\ell,k)$ be the array obtained by deleting the $\ell$-th and the $k$-th columns of $A(\bfx)$ for $1\leq \ell\ne k\le n/b$. It is clear that $A(\ell,k)\in\cA_2^b(\bfx)$ and
  $$
  \cup_{1\le\ell\ne k\le\frac{n}{b}}\mathset{A(\ell,k)_i} = \cB_2(A(\bfx)_i),
  $$
  for all $1\leq i \leq b$.
By \cite[eq. (11)]{Levenshtein2001JCTA}, we know that
$$
\binom{r-1}{2}\le\abs{\cB_2(\bbv)}\le\binom{r+1}{2}
$$
for any sequence $\bbv\in\Sigma_q^n$ with exactly $r$ runs.
Since $\abs{\cA_2^b(\bfx)} \geq \max_{1\leq i \leq b} \abs{ \cB_2 (A(\bfx)_i) }$ and $\abs{\cB_2^b(\bfx)}=\abs{\cA_2^b(\bfx)}$ for all $\bfx$, it follows that
\begin{equation}\label{eq_upperbound1}
\begin{aligned}
    \abs{\cB_2^b(\bfx)}&\ge\max_{1\leq i \leq b} \abs{\cB_2 (A(\bfx)_i)}\\
    &\ge\binom{\max_{1\le i\le b}\mathset{r\parenv{A(\bfx)_i}}-1}{2}\ge \binom{t+1}{2},
\end{aligned}
\end{equation}
for all $\bfx\in\cC_1$. Since $\cC$ is a $2$-$b$-burst-deletion correcting code, we know that $\cC_1$ is also a $2$-$b$-burst-deletion correcting code. So we have $\cB_2^b(\bfx)\cap\cB_2^b(\bby)=\emptyset$ for all distinct $\bfx,\bby\in\cC_1$. Then it follows from \Cref{eq_upperbound1} that $\abs{\cC_1}\binom{t+1}{2}\le\sum_{\bfx\in\cC_1}\abs{\cB_2^b(\bfx)}\le q^{n-2b}$ and hence 
\begin{equation}\label{eq_upperbound2}
\begin{aligned}
    \abs{\cC_1}\le\frac{q^{n-2b}}{\binom{t+1}{2}}&\le \frac{q^n}{n^2q^{2b}}\cdot\frac{2n^2}{t^2}\\
    &=\frac{q^n}{n^2q^{2b}}\cdot\frac{2n^2}{\parenv{1-1/q-\epsilon}^2\parenv{n/b-1}^2}\\
    &=\frac{q^n}{n^2q^{2b}}\cdot\frac{2b^2}{(1-1/q-\epsilon)^2(1-b/n)^2}\\
    &=\frac{2b^2q^{n+2}}{n^2q^{2b}(q-1)^2}\cdot\frac{1}{\parenv{1-\frac{q}{q-1}\epsilon}^2\parenv{1-b/n}^2}\\
    &\le \frac{3b^2q^{n+2}}{n^2q^{2b}(q-1)^2}
\end{aligned}
\end{equation}
as long as $\parenv{1-\frac{q}{q-1}\epsilon}^2\parenv{1-b/n}^2\ge 2/3$, which is possible when $n$ is sufficiently large.

Next, we proceed to upper bound $\abs{\cC_2}$. To that end, define $\cC^\prime$ to be the following set
$$
\mathset{\bfx=\bfx^{(1)}\cdots\bfx^{(b)}\in\Sigma_q^n: 
\begin{array}{c}
  \bfx^{(i)}\in\Sigma_q^{n/b},\; \forall 1\le i\le b  \\
     r\parenv{\bfx^{(i)}}\le t+1,\; \forall 1\le i\le b
\end{array}
}.
$$
Since $r\parenv{A(\bfx)_i}\le t+1$ for any $\bfx\in\cC_2$ and $1\le i\le b$, we conclude that $\abs{\cC_2}\le\abs{\cC^\prime}$. So it suffices to estimate an upper bound of $\abs{\cC^\prime}$.

Since $m = n/b-1$, by definition of $\cC^\prime$, we have
\begin{equation}\label{eq_upperbound3}
\begin{aligned}
    \abs{\cC^\prime}&=\abs{\mathset{\bbv \in \Sigma_q^{m+1}: r(\bbv) \leq t+1}}^b\\
    &=\abs{\cup_{j=1}^{t+1} \mathset{\bbv \in \Sigma_q^{m+1}: r(\bbv) = j}}^b\\
    &=\parenv{\sum_{j=1}^{t+1}\abs{\mathset{\bbv \in \Sigma_q^{m+1}: r(\bbv) = j}}}^b\\
    &\overset{(a)}{=}\parenv{q\sum_{j=1}^{t+1} \binom{m}{j-1}(q-1)^{j-1}}^b\\
    &=\parenv{q\sum_{j=0}^{t} \binom{m}{j}(q-1)^{j}}^b
\end{aligned}
\end{equation}
where (a) follows from the well-known result (see the proof of \cite[Theorem 3.1]{Kulkarni2013IT}):
$$\abs{\mathset{\bbv \in \Sigma_q^{m+1}: r(\bbv) = j}} =\binom{m}{j-1}q (q-1)^{j-1}.$$

Since $t=(1-1/q-\epsilon)m$, we have $t/m=1-1/q-\epsilon\le 1-1/q$. Then it follows from \cite[Proposition 3.3.3]{Guruswami2023} that
\begin{equation}\label{eq_upperbound4}
\sum_{j=0}^{t} \binom{m}{j}(q-1)^{j} \leq q^{mH_q(t/m)}=q^{mH_q(1-1/q-\epsilon)},
\end{equation}
where $H_q(x)\triangleq x \log_q(q-1) -x\log_q x - (1-x)\log_q (1-x)$ is the $q$-ary entropy function. Here, $\log_q x=\frac{\log x}{\log q}$ for any positive real number $x$.

If $\frac{\log n}{n}<\frac{\log q}{12}f(q)^2$, then $\epsilon<f(q)$. Now by \Cref{lem_A1}, we have
$$
H_q(1 - \frac{1}{q} - \epsilon)\leq 1 - \frac{\epsilon^2}{4}.
$$
Now it follows from \Cref{eq_upperbound3,eq_upperbound4} that
\begin{equation}\label{eq_upperbound5}
\begin{aligned}
  \abs{\cC_2}\le\abs{\cC^\prime}\leq q^{b +(n-b)(1- \epsilon^2/4)}&\overset{(a)}{=}q^{b +(n-b)\parenv{1-\frac{3\log n}{n\log q}}}\\
  &=\frac{q^n}{n^3}\parenv{\sqrt[n]{n}}^{3b}\\
  &\overset{(b)}{<}(1.121)^{3b}\frac{q^n}{n^3},
\end{aligned}
\end{equation}
where (a) follows from selecting $\epsilon=\sqrt{\frac{12\log n}{n\log q}}$ and (b) follows from the fact that $\sqrt[n]{n}<1.121$ when $n\ge 30$. 
We conclude the proof by combining \Cref{eq_upperbound2,eq_upperbound5}.
\end{IEEEproof}

Then next corollary is a direct consequence of \Cref{thm_lowerbound,thm_upperbound}.
\begin{corollary}
    Suppose that $q$ and $b$ are fixed. Let $f(q)$ be defined in \Cref{thm_upperbound}. When $n\ge\max\mathset{30,\frac{(1.21)^{3b}q^{2b-2}(q-1)^2}{b^2}}$ is sufficiently large such that $\frac{\log n}{n}<\frac{\log q}{12}f(q)^2$ and $\parenv{1-\frac{q}{q-1}\sqrt{\frac{12\log n}{n\log q}}}^2\parenv{1-b/n}^2\ge 2/3$, we have
    $$
    2\log n+\log\parenv{\frac{q^{2b-2}(q-1)^2}{4b^2}}\le\rho(\cC)\le 4\log n+2b\log q,
    $$
    where $\cC\subseteq\Sigma_q^n$ is a $2$-$b$-burst-deletion correcting code with maximum size.
\end{corollary}
\begin{IEEEproof}
   Firstly, the upper bound follows straightforward from \Cref{thm_lowerbound} just by noticing that $\binom{n}{2}^2\le n^4$. Since $n\ge \frac{(1.21)^{3b}q^{2b-2}(q-1)^2}{b^2}$, we have $\parenv{\frac{3b^2}{q^{2b-2}(q-1)^2}+\frac{(1.121)^{3b}}{n}}\frac{q^n}{n^2}\le \frac{4b^2}{q^{2b-2}(q-1)^2}\frac{q^n}{n^2}$. Now the lower bound follows from \Cref{thm_upperbound}.
\end{IEEEproof}

\section{Codes For Correcting Two $b$-Burst-Deletions}\label{sec_qburst}
%%%%%%%%%%%%%%%%%%%%%%%%%%%%%%%%%%%%%%%%%%%%%%%%%%
In this section, we construct $q$-ary $2$-$b$-burst-deletion correcting codes with redundancy at most $5\log n+O(\log\log n)$, for any $q\ge2$ and $b>1$. Our idea is to first locate positions of deletions in short intervals (which is accomplished in \Cref{sec_errorposition}, \Cref{lem_qpositions}) and then correct errors in these intervals (which is accomplished in \Cref{sec_qruns,sec_qinterval}). 

\subsection{Approximately determine positions of deletions}\label{sec_errorposition}
%%%%%%%%%%%%%%%%%%%%%%%%%%%%%%%%%%%%%%
\begin{definition}[regularity]
    A sequence $\bfx\in\Sigma_2^n$ is said to be $d$-\textbf{regular} if each substring of $\bfx$ of length at least $d\log n$ contains both $00$ and $11$. 
\end{definition}

For the number of regular sequences, we have the following lemma.
\begin{lemma}\cite[Lemma 11]{Guruswami2021it}\label{lem_numberregular}
The number of $d$-regular sequences of length $n$ is at least $2^{n-1}$, as long as $d\ge 7$ and $n$ be such that $\floorenv{\frac{d}{2}\log n}n^{0.15d-1}\ge 12$. In particular, when $d=7$, it suffices to require $n\ge 9$.
\end{lemma}

The next lemma ensures that we can efficiently correct two deletions in $d$-regular sequences.
\begin{lemma}\cite[Theorem 7]{Guruswami2021it}\label{lem_rtwodeletion}
Suppose $d\ge 7$. There exists a function $\eta:\Sigma_2^n\rightarrow\Sigma_2^{4\log n+10\log\log n+O_d(1)}$, computable in linear time, such that for any $d$-regular sequence $\bfx\in\Sigma_2^n$, given $\eta\parenv{\bfx}$ and any $\bfx^\prime\in\cB_2\parenv{\bfx}$, one can uniquely and efficiently recover $\bfx$. Here, $O_d(1)$ denotes a constant depending only on $d$.
\end{lemma}

Furthermore, once $\bfx$ is recovered, locations of the two deletions in $\bfx$ can be approximated, as ensured by the following lemma.
\begin{lemma}\cite[Lemma 9]{Song2023IT}\label{lem_positions}
    Suppose that $\bfx\in\Sigma_2^n$ is $d$-regular and $\bfx^{\prime}\in\Sigma_2^{n-2}$ is obtained from $\bfx$ by deleting two symbols $x_{i_1}$ and $x_{i_2}$. When given $\bfx$ and $\bfx^\prime$, we can
    \begin{enumerate}[$(1)$]
        \item either find distinct runs $\bfx_{J_1}$ and $\bfx_{J_2}$ of $\bfx$, uniquely determined by $\bfx$ and $\bfx^{\prime}$, such that $i_1\in J_1$ and $i_2\in J_2$.
        \item or find an interval $J\subseteq[n]$, uniquely determined by $\bfx$ and $\bfx^\prime$, of length at most $3d\log n$ such that $i_1,i_2\in J$.
    \end{enumerate}
\end{lemma}
A proof of \Cref{lem_positions} is given in \cite{Song2023IT}. Here we show the intuition behind this lemma.
\begin{example}
    \begin{enumerate}[$(1)$]
        \item Let $\bfx=000111$ and $\bfx^\prime=0011$. Then $\bfx^\prime$ is obtained from $\bfx$ by deleting one bit in the run $000$ and one bit in the run $111$. This corresponds to case (1) in \Cref{lem_positions}.
        \item $\bfx=10010101110$ and $\bfx^\prime=100101110$. Then $\bfx^\prime$ can be obtained from $\bfx$ by deleting $x_2$ and $x_4$, or by deleting $x_3$ and $x_4$, or by deleting $x_4$ and $x_5$, or by deleting $x_7$ and $x_8$, or by deleting $x_7$ and $x_9$. Therefore, we can only locate error positions in the substring $001010111$, which is the concatenation of a run $00$, an alternating substring $1010$ and a run $111$. If $\bfx\in\Sigma_2^n$ is $d$-regular, then each run has length at most $d\log n$ and each alternating substring has length at most $d\log n$. This is why in case (2) of \Cref{lem_positions} we can locate error positions in an interval of length at most $3d\log n$.
    \end{enumerate}
\end{example}

Now we briefly explain why \Cref{lem_positions} is helpful. Suppose $\bfx\in\Sigma_2^n$ and $\bfx^\prime\in\cB_2^b(\bfx)$. Let $A(\bfx)=A(\bfx,b)$ and $A(\bfx^\prime)=A(\bfx^\prime,b)$ be the matrix representations of $\bfx$ and $\bfx^\prime$, respectively. If $A(\bfx)_1$ is $d$-regular (where $d\ge7$) and $\eta(A(\bfx)_1)$ is given, \Cref{lem_rtwodeletion} ensures that we can decode $A(\bfx)_1$ from $A(\bfx^\prime)_1$ and \Cref{lem_positions} ensures that error positions in $A(\bfx)_1$ can be approximately determined in one or two short intervals. By \Cref{obs_deletionposition}, this further reveals information of error positions in remaining rows of $A(\bfx)$, or equivalently, error positions in $\bfx$. Detailed analysis will be given \Cref{lem_qpositions}. Before that, it is convenient to generalize the notion of regularity to general alphabets. To that end, we associate with any $q$-ary sequence a binary sequence.

Let $q\ge2$. For any $x\in\Sigma_q$, we can uniquely write it as $x=\ceilenv{q/2}u_x+v_x$, where $u_x\in\{0,1\}$ and $0\le v_x<\ceilenv{q/2}$. Then for a sequence $\bfx\in\Sigma_q^n$, define $u(\bfx)\triangleq u_{x_1}\cdots u_{x_n}$, i.e., $u(\bfx)_i=u_{x_i}$ for all $1\le i\le n$. We call $u(\bfx)$ the binary sequence associated with $\bfx$. Clearly, $u(\bfx)\in\Sigma_2^n$ and when $q=2$, we have $u(\bfx)=\bfx$. It should be noted that this decomposition of $x\in\Sigma_q$ into $u_x$ and $v_x$ was also applied in \cite[Section V]{Tuan2024IT} to construct a single-burst-deletion of variable length.
\begin{example}\label{example_decomposition}
    Let $q=3$. We have that $\ceilenv{q/2}=2$. If $x=1$, then $u_x=0$ and $v_x=1$. If $x=2$, then $u_1=1$ and $v_x=0$.
\end{example}

\begin{remark}
Although any $x\in\Sigma_q$ can be decomposed as $x=\ceilenv{q/2}u_x+v_x$, it might exist some $u\in\{0,1\}$ and $0\le v<\ceilenv{q/2}$ such that $\ceilenv{q/2}u+v\notin\Sigma_q$. In fact, if $q$ is even, then for any $u\in\{0,1\}$ and any $0\le v<\ceilenv{q/2}$, we have $\ceilenv{q/2}u+v\in\Sigma_q$. But when $q$ is odd, we have $\ceilenv{q/2}+\ceilenv{q/2}-1=q\notin\Sigma_q$.
\end{remark}

\begin{definition}
    A sequence $\bfx\in\Sigma_q^n$ is said to be $d$-regular if $u(\bfx)$ is $d$-regular. In other words, $\bfx$ is $d$-regular if and only if any substring of $\bfx$ of length at least $d\log n$ contains two consecutive coordinates smaller than $\ceilenv{q/2}$ and two consecutive coordinates no less than $\ceilenv{q/2}$.
\end{definition}

Let $\cR_{q,n,d}\triangleq\mathset{\bfx\in\Sigma_q^n:\bfx\text{ is }d\text{-regular}}$. To estimate a lower bound on our code size in \Cref{thm_qtwoburst}, we need a lower bound on $\abs{\cR_{q,n,d}}$.
\begin{lemma}\label{lem_qregular}
    Let $q\ge 2$. 
    \begin{itemize}
        \item  If $q$ is even, then $\abs{\cR_{q,n,d}}\ge q^{n}/2\ge q^{n-1}$ for all $d\ge 7$ and $n$ such that $\floorenv{\frac{d}{2}\log n}n^{0.15d-1}\ge 12$. In particular, when $d=7$, it suffices to require $n\ge 9$.
        \item If $q$ is odd, then $\abs{\cR_{q,n,d}}\ge q^{n-1}$ for all $d\ge 10$ and $\floorenv{\frac{d}{2}\log n}n^{-1-\frac{d}{2}\log(0.87)}\ge \frac{200q}{87(q-1)}$. Note that when $d\ge 10$, we have $-1-\frac{d}{2}\log(0.87)>0$ and hence this condition could be satisfied when $n$ is sufficiently large. In particular, when $d=10$, it suffices to require $\floorenv{5\log n}n^{-1-5\log(0.87)}\ge \frac{200q}{87(q-1)}$.
    \end{itemize}
\end{lemma}
\begin{IEEEproof}
 Suppose first that $q$ is even. Any $x\in\Sigma_q$ can be uniquely represented as  $x=\frac{q}{2} u_x+v_x$, where $u_x\in\{0,1\}$ and $0\le v_x<q/2$. On the other hand, when $q$ is even, we have $\frac{q}{2} u_x+v_x\in\Sigma_q$ for any $u_x\in\{0,1\}$ and $0\le v_x<q/2$. Then it follows from \Cref{lem_numberregular} that $\abs{\cR_{q,n,d}}\ge\abs{\cR_{2,n,d}}(q/2)^n=q^n/2\ge q^{n-1}$.

 Now suppose that $q\ge 3$ is odd. In this case, the previous argument does not hold. This is because when $q$ is odd, $u_x=1$ and $v_x=(q-1)/2$, we have $\ceilenv{q/2}u_x+(q-1)/2=q\notin\Sigma_q$. Fortunately, we can apply similar ideas in proofs of \cite[Lemma 11]{Guruswami2021it} and \cite[Lemma 5]{ShuLiu2024IT} to derive our lower bound on $\abs{\cR_{q,n,d}}$. For $m\ge 2$, define
 \begin{align*}
     S_m^L=\mathset{\bbv\in\Sigma_q^m:
     \begin{array}{c}
     \nexists 1\le i< m\text{ such that}\\
     0\le v_i,v_{i+1}\le(q-1)/2
     \end{array}
     },\\
     S_m^U=\mathset{\bbv\in\Sigma_q^m:
     \begin{array}{c}
     \nexists 1\le i< m\text{ such that}\\
     (q+1)/2\le v_i,v_{i+1}<q
     \end{array}
     }.
 \end{align*}
 
 We next prove by induction on $m$ that $\abs{S_m^L}<(0.87q)^m$ for all $m\ge2$. By the inclusion-exclusion principle, it is easy to verify that $\abs{S_2^L}=q^2-\parenv{\frac{q+1}{2}}^2=\frac{3}{4}q^2-\frac{1}{2}q-\frac{1}{4}<0.75q^2<(0.87q)^2$ and $\abs{S_3^L}=q^3-2q\parenv{\frac{q+1}{2}}^2+\parenv{\frac{q+1}{2}}^3=\frac{5}{8}q^3-\frac{5}{8}q^2-\frac{1}{8}q+\frac{1}{8}<0.625q^3<(0.87q)^3$. Now suppose that $m\ge4$ and the conclusion is proved for all $m^\prime\le m-2$. For $\bbv\in S_m^L$, it must hold that $v_1v_2\in S_2^L$ and $v_3\cdots v_m\in S_{m-2}^L$. So it follows that $\abs{S_m^L}\le\abs{S_2^L}\abs{S_{m-2}^L}<(0.87q)^2(0.87q)^{m-2}=(0.87q)^m$.

 For $\bbv\in\Sigma_q^m$, let $\overline{\bbv}\triangleq (q-v_1)\cdots(q-v_m)$. It is easy to verify that if $\bbv\in S_m^U$, then $\overline{\bbv}\in S_m^L$. Then it follows that $\abs{S_m^U}\le\abs{S_m^L}<(0.87q)^m$.

 Let $m=\floorenv{\frac{d}{2}\log n}$ and 
 $$
 Q=\mathset{\bbv\in\Sigma_q^m:
 \begin{array}{c}
 \exists i,j\text{ such that }v_i,v_{i+1}<(q+1)/2\\
 \text{and }v_j.v_{j+1}\ge(q+1)/2
 \end{array}
 }.
 $$
 In other words, $Q$ is the set of length-$m$ sequences which contain two consecutive coordinates smaller than $\ceilenv{q/2}$ and two consecutive coordinates no less than $\ceilenv{q/2}$. Then we have $\abs{Q}\ge q^m-\abs{S_m^L}-\abs{S_m^U}>q^m-2(0.87q)^m$. Now let $k=\floorenv{n/m}$ and define $\cR^\prime$ to be the following set
 $$
 \mathset{\bfx=\bfx^{(1)}\cdots\bfx^{(k)}\bfx^{(k+1)}\in\Sigma_q^n:
 \begin{array}{c}
    \bfx^{(i)}\in Q,\forall 1\le i\le k  \\
     \bfx^{(k+1)}\in\Sigma_q^{n-km}
 \end{array}
 }.
 $$
It is easy to see that for any $\bfx\in\cR^\prime$, any substring of $\bfx$ of length at least $d\log n$ must contain some $\bfx^{(i)}$, where $1\le i\le k$. Therefore, $\bfx$ is $d$-regular and hence $\cR^\prime\subseteq\cR_{q,n,d}$. This implies that
 \begin{align*}
     \abs{\cR_{q,n,d}}\ge\abs{\cR^\prime}&=\abs{Q}^kq^{n-km}\\
     &>\parenv{q^m-2(0.87q)^m}^kq^{n-km}\\
     &=q^n\parenv{1-2(0.87)^m}^k\\
     &\overset{(a)}{\ge}q^n\parenv{1-2k(0.87)^m}\\
     &=q^n\parenv{1-2\floorenv{\frac{n}{m}}2^{\floorenv{\frac{d}{2}\log n}\log(0.87)}}\\
     &\ge q^n\parenv{1-2\frac{n}{m}2^{\floorenv{\frac{d}{2}\log n}\log(0.87)}}\\
     &\overset{(b)}{\ge} q^n\parenv{1-2\frac{n}{m}2^{\frac{d}{2}\log n\cdot\log(0.87)-\log(0.87)}}\\
     &=q^n\parenv{1-\frac{2}{0.87\floorenv{\frac{d}{2}\log n}}n^{1+\frac{d}{2}\log(0.87)}}\\
     &\overset{(c)}{\ge} q^{n-1}.
 \end{align*}
 Here, (a) follows from the fact that $(1+x)^r\ge 1+rx$ for any integer $r\ge 1$ and any real number $x\ge -1$; (b) follows from the fact that $\log(0.87)<0$ and $\floorenv{\frac{d}{2}\log n}\ge\frac{d}{2}\log n-1$; (c) follow from the fact that $d\ge 10$ and $\floorenv{\frac{d}{2}\log n}n^{-1-\frac{d}{2}\log(0.87)}\ge \frac{200q}{87(q-1)}$.
\end{IEEEproof}

In Appendix~\ref{appendix2}, we will discuss how to encode a sequence into a $d$-regular sequence.

Now we return to the aim of this subsection: approximately determine error positions. Let $\bfx\in\Sigma_q^n$ and $\bfx^\prime$ be obtained from $\bfx$ by a $2$-$b$-burst-deletion.
Let $u(\bfx)$ and $u(\bfx^\prime)$ be the binary sequences associated with $\bfx$ and $\bfx^\prime$, respectively (see the paragraph prior to \Cref{example_decomposition}). Then we have $u(\bfx^\prime)\in\cB_2^b\parenv{u(\bfx)}$. By the relationship between $\bfx$ and $u(\bfx)$, it suffices to locate the $2$-$b$-burst-deletion in $u(\bfx)$. In the rest of this section, let $U(\bfx)=A(u(\bfx),b)$ be the matrix representation of $u(\bfx)$. Denote the $i$-th row of $U(\bfx)$ by $U(\bfx)_i$. Similarly, we can define $U(\bfx^\prime)$ and $U(\bfx^\prime)_i$.

Since $u(\bfx^\prime)\in\cB_2^b\parenv{u(\bfx)}$, it follows from \Cref{obs_deletionposition} that $U(\bfx^\prime)_i$ is obtained from $U(\bfx)_i$ by two deletions. Suppose that $U(\bfx)_1$ is $d$-regular and $\eta\parenv{U(\bfx)_1}$ is given. According to \Cref{lem_rtwodeletion}, we can efficiently recover $U(\bfx)_1$. Then by \Cref{lem_positions}, the two deletions in $U(\bfx)_1$ can be approximately located. In \Cref{lem_qpositions} below, we will show that this helps to locate errors in $u(\bfx)$. To explain the idea, more notations are needed.

For $1\le i\le b$ and $1\le j\le n/b$, let $U(\bfx)_{i,j}$ be the $j$-th coordinate of $U(\bfx)_i$. By the relationship between $u(\bfx)$ and $U(\bfx)$, it is easy to see that $U(\bfx)_{1,j}=u(\bfx)_{1+(j-1)b}$. This justifies the following definition
\begin{equation*}
    I_{n,b}\triangleq\mathset{1+(j-1)b: 1\le j\le n/b}.
\end{equation*}
Note that $I_{n,b}$ is a subset of $[n]$. Clearly, we have $U(\bfx)_1=u(\bfx)_{I_{n,b}}$.
\begin{example}\label{examp_location1}
    Let $n=10$, $b=2$ and $\bfx=\rcomment{0}1\rcomment{2}2\rcomment{2}1\rcomment{0}1\rcomment{1}2\in\Sigma_3^{10}$. Then $u(\bfx)=\rcomment{0}0\rcomment{1}1\rcomment{1}0\rcomment{0}0\rcomment{0}1$. By definition, we have $I_{n,b}=\{1,3,5,7,9\}$ and $u(\bfx)_{I_{n,b}}=01100$. On the other hand, the matrix representations of $u(\bfx)$ is 
    $$
    U(\bfx)=
    \begin{pmatrix}
        0&1&1&0&0\\
        0&1&0&0&1
    \end{pmatrix}.
    $$
    It is easy to see that $U(\bfx)_1=u(\bfx)_{I_{n,b}}$.
\end{example}

Recall that $r\parenv{U(\bfx)_1}$ is the number of runs in $U(\bfx)_1$. For $1\le j\le r\parenv{U(\bfx)_1}$, let $U(\bfx)_{1,I_j}$ be the $j$-th run of $U(\bfx)_1$. Furthermore, suppose $I_j=\sparenv{p_{j-1}+1,p_{j}}$, where $p_0=0$, $p_{r\parenv{U(\bfx)_1}}=n/b$ and $p_{j-1}<p_j$ for all $j$.
By definition, $I_j$ is an interval in $[n/b]$. For all $j\ge 1$, we associate with $I_j$ an interval in $[n]$: $I_j^\prime\triangleq\sparenv{p_{j-1}b+1,p_{j}b}$. In addition, define $I_j^L\triangleq\sparenv{(p_{j-1}-1)b+2,p_{j-1}b}\cap[n]$. In other words, $I_j^L$ is the interval of length at most $b-1$ in $[n]$ to the left of $I_j^\prime$. Note that $I_j$, $I_j^\prime$ and $I_j^L$ are dependent on $\bfx$. We omit $\bfx$ for simpler notations. 
\begin{example}
    Let $\bfx$ be the sequence in \Cref{examp_location1}. There are three runs in $U(\bfx)_1=01100$: $0$, $11$ and $00$. Recall that $p_j$ indicates where the $j$-th run ends. Then we have $p_1=1$, $p_2=3$, $p_3=5$, $I_1=[1,1]$, $I_2=[2,3]$ and $I_3=[4,5]$. All intervals $I_j^\prime$ and $I_j^L$ are listed in the following:
    $$
    \begin{array}{cc}
         I_1^L=\emptyset, I_1^\prime=[1,2], \\
        I_2^L=[2,2], I_2^\prime=[3,6],\\
        I_3^L=[6,6], I_3^\prime=[7,10].
    \end{array}
    $$
\end{example}

\begin{figure}[t]
  \centering
  {
  \subfigure[$U(\bfx)$: $I_j$ in $\sparenv{n/b}$]
  {
  \begin{minipage}[H]{0.45\textwidth}
  \centering
    \begin{tikzpicture}
     \path (-4.6,1.9)node{\tiny{the first row}};
     \path (-4,1.9)node{\tiny{$\rightarrow$}};
     \draw[dotted,thick] (-4.3,1)--(-3.85,1);
     \draw (-3.75,0)rectangle(-3.25,2);
     \draw (-3.75,1.8)rectangle(-3.25,2);
     \filldraw[fill=green!20] (-3.75,0)rectangle(-3.25,1.8);
     \draw (-3,0)rectangle(-2.5,2);
     \draw[shade] (-3,1.8)rectangle(-2.5,2);
     \draw[dotted,thick] (-2.45,1)--(-1.45,1);
     \draw (-1.4,0)rectangle(-0.9,2);
     \draw[shade] (-1.4,1.8)rectangle(-0.9,2);
     \filldraw[fill=red] (-1.4,0)rectangle(-0.9,0.2);
     \draw[dotted,thick] (-0.85,1)--(-0.4,1);
     \path (-3.5,2.3)node{\tiny{$p_{j-1}$}};
     \path (-3.5,2.1)node{\tiny{$\downarrow$}};
     \path (-2.7,2.4)node{\tiny{$p_{j-1}+1$}};
     \path (-2.7,2.2)node{\tiny{$\downarrow$}};
     \path (-1.15,2.4)node{\tiny{$p_{j}$}};
     \path (-1.15,2.2)node{\tiny{$\downarrow$}};
     \path (-2.3,1.9)node{\tiny{$\leftarrow$}};
     \path (-1.6,1.9)node{\tiny{$\rightarrow$}};
     \path (-1.95,1.9)node{\tiny{$I_{j}$}};
     \draw[decorate,decoration={brace,raise=8pt}] (-0.6,2)--(-0.6,0);
     \path (0.05,1)node{\tiny{$b$ rows}};
    \end{tikzpicture}
  \end{minipage}
  }
  \subfigure[$\bfx$: $I_{j}^\prime$ and $I_j^L$ in $\sparenv{n}$]
  {
  \begin{minipage}[H]{0.45\textwidth}
  \centering
    \begin{tikzpicture}
     \draw[dotted,thick](0,0.25)--(0.35,0.25);
     \filldraw[fill=green!20] (0.4,0)rectangle(2,0.5);
     \draw (2,0)rectangle(3.8,0.5);
     \draw[shade] (2,0)rectangle(2.2,0.5);
     \path (2.1,0.9)node{\tiny{$p_{j-1}b+1$}};
     \path (2.1,0.7)node{\tiny{$\downarrow$}};
     \draw[dotted,thick](3.85,0.25)--(4.65,0.25);
     \draw (4.7,0)rectangle(6.5,0.5);
     \draw[shade] (4.7,0)rectangle(4.9,0.5);
     \filldraw[fill=red] (6.3,0)rectangle(6.5,0.5);
     \path (6.4,0.9)node{\tiny{$p_{j}b$}};
     \path (6.4,0.7)node{\tiny{$\downarrow$}};
     \draw[dotted,thick](6.55,0.25)--(6.9,0.25);
     \draw[decorate,decoration={brace,raise=8pt}] (6.5,0.2)--(2,0.2);
     \path (4.25,-0.5)node{\tiny{$I_{j}^\prime$}};
     \draw[decorate,decoration={brace,raise=8pt}] (2,0.2)--(0.4,0.2);
     \path (1.2,-0.5)node{\tiny{$I_{j}^L$}};
    \end{tikzpicture}
  \end{minipage}
  }
  }
  \caption{Illustration of relationship between $I_j$, $I_{j}^\prime$ and $I_j^L$.}\label{fig_threeinterval}
\end{figure}
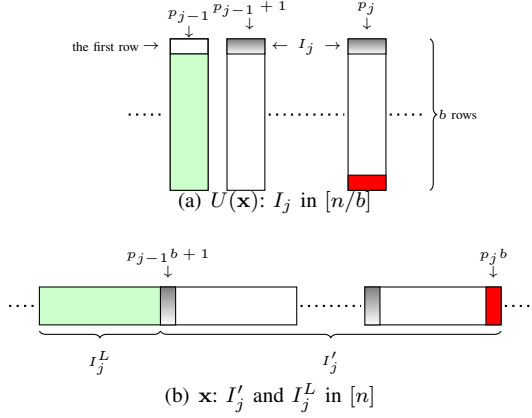

We briefly explain why we define intervals $I_j^\prime$ and $I_j^L$. A $b$-burst-deletion in $u(\bfx)$ induces one deletion in each row of $U(\bfx)$. If it is known that the deletion in the first row $U(\bfx)_1$ occurs in the interval $I_j=\sparenv{p_{j-1}+1,p_j}$, we can only locate the deletion in the $i$-th row in the interval $\sparenv{p_{j-1},p_j}$ when $i\ge2$ (see \Cref{obs_deletionposition}). Note that $U(\bfx)_1=u(\bfx)_{I_{n,b}}$. When looking at $u(\bfx)$, interval $I_j^L$ (if not empty) corresponds to the last $b-1$ coordinates in the $p_{j-1}$-th column of $U(\bfx)$ and interval $I_j^\prime$ corresponds to columns of $U(\bfx)$ indexed by $I_j$. As a result, the $b$-burst-deletion in $u(\bfx)$ can be located in the interval $I_j^L\cup I_j^\prime$. See \Cref{fig_threeinterval} for an illustration.

In the rest of this section, let $A(\bfx)=A(\bfx,b)$ be the matrix representation of $\bfx$ and $A(\bfx)_1$ be the first row of $A(\bfx)$.
\begin{lemma}\label{lem_qpositions}
    Let $q\ge2$, $b>1$ and $n>2b$. Suppose $\bfx\in\Sigma_q^n$ and $\bfx^\prime=\bfx_{[n]\setminus(D_1\cup D_2)}$, where $D_1,D_2\subseteq[n]$ are two disjoint intervals of length $b$. That is to say, $\bfx^\prime$ is obtained from $\bfx$ by a $2$-$b$-burst-deletion. Suppose that $A(\bfx)_1$ is $d$-regular, that is, $U(\bfx)_1$ is $d$-regular. Let $\eta(\cdot)$ be the function defined in \Cref{lem_rtwodeletion}. Then given $\eta\parenv{U(\bfx)_1}$, we can
    \begin{enumerate}[$(1)$]
        \item either find $1\le j_1<j_2\le r\parenv{U(\bfx)_1}$, such that
        $
        D_1\subseteq I_{j_1}^L\cup I_{j_1}^\prime$ and $D_2\subseteq I_{j_2}^L\cup I_{j_2}^\prime$;
        \item or find an interval $J\subseteq[n]$ of length at most $3bd\log(n/b)+b-1$, such that $D_1,D_2\subseteq J$.
    \end{enumerate}
\end{lemma}
\begin{IEEEproof}
   Since $\bfx^\prime=\bfx_{[n]\setminus(D_1\cup D_2)}$, we have $u(\bfx^\prime)=u(\bfx)_{[n]\setminus(D_1\cup D_2)}$. It follows that $U(\bfx^\prime)_1\in\cB_2\parenv{U(\bfx)_1}$.
    Suppose that $U(\bfx^\prime)_1$ is obtained from $U(\bfx)_1$ by deleting $U(\bfx)_{1,k_1}$ and $U(\bfx)_{1,k_2}$. Recall that $U(\bfx)_1=u(\bfx)_{I_{n,b}}$. This implies that $U(\bfx^\prime)_1$ is obtained from $u(\bfx)_{I_{n,b}}$ by deleting $u(\bfx)_{1+(k_1-1)b}$ and $u(\bfx)_{1+(k_2-1)b}$.
    
    Since $U(\bfx)_1$ is $d$-regular, according to \Cref{lem_rtwodeletion}, we can recover $U(\bfx)_1$  from $U(\bfx^\prime)_1$ with the help of $\eta\parenv{U(\bfx)_1}$. Then we can know the number of runs in $U(\bfx)_1$ and $I_j$ for all $1\le j\le r\parenv{U(\bfx)_1}$. In addition, by \Cref{lem_positions}, we can
    \begin{itemize}
        \item either find $1\le j_1<j_2\le r\parenv{U(\bfx)_1}$, such that $k_1\in I_{j_1}$ and $k_2\in I_{j_2}$;
        \item or find an interval $[c,d]\subseteq[n/b]$ of length at most $3d\log(n/b)$ such that $k_1,k_2\in[c,d]$.
    \end{itemize}
   Similar to the discussion in the paragraph above \Cref{lem_qpositions}, for the first case, we have $D_1\subseteq \sparenv{p_{j_1-1}b-b+2,p_{j_1}b}=I_{j_1}^L\cup I_{j_1}^\prime$ and $D_2\subseteq \sparenv{p_{j_2-1}b-b+2,p_{j_2}b}=I_{j_2}^L\cup I_{j_2}^\prime$. For the second case, we have $D_1,D_2\subseteq J\triangleq[(c-2)b+2,db]\cap[n]$. Now the proof is completed.
\end{IEEEproof}

From now on, we assume that $\bfx\in\Sigma_q^n$ and $A(\bfx)_1$ is $d$-regular, where $d$ will be specified later. Let $\bfx^{\prime}=\bfx_{[n]\setminus(D_1\cup D_2)}\in\cB_2^b\parenv{\bfx}$, where $D_1,D_2\subseteq[n]$ are two unknown disjoint intervals of length $b$. Suppose that $\eta\parenv{U(\bfx)_1}$ is given. The first step of decoding $\bfx$ from $\bfx^\prime$ is to decode $U(\bfx)_1$ from $U(\bfx^\prime)_1$. After this step, we can know the number of runs in $U(\bfx)_1$ and $I_j$ for all $1\le j\le r\parenv{U(\bfx)_1}$. Then by \Cref{lem_qpositions}, one can approximately locate the two bursts in $\bfx$ if $\eta\parenv{U(\bfx)_1}$ is given. Based on the two cases in \Cref{lem_qpositions}, we split our discussion into \Cref{sec_qruns} and \Cref{sec_qinterval}. Our codes for correcting two $b$-burst-deletions are presented in \Cref{thm_qtwoburst}, whose proof is divided into \Cref{lem_qruns,lem_qinterval}. In \Cref{sec_qruns} and \Cref{sec_qinterval}, we follow notations defined in this subsection.

Before proceeding, we need the following observation.
\begin{observation}\label{obs_singleburst}
Let $m>b$. Suppose $\bbu\in\Sigma_q^m$ and $\bbv$ is obtained from $\bbu$ by single $b$-burst-deletion. Then for any $b\le i\le m-b$, we have $\bbv_{[1,i-b]}\in\cB_1^b\parenv{\bbu_{[1,i]}}$ and $\bbv_{[i+1,m-b]}\in\cB_1^b\parenv{\bbu_{[i+1,m]}}$.
\end{observation}

\subsection{When Case (1) in \Cref{lem_qpositions} happens}\label{sec_qruns}
%%%%%%%%%%%%%%%%%%%%%%%%%%%%%%%%%%%%%%%%%%%%
In this case, we already found $1\le j_1<j_2\le r\parenv{U(\bfx)_1}$, such that $D_1\subseteq I_{j_1}^L\cup I_{j_1}^\prime$ and $D_2\subseteq I_{j_2}^L\cup I_{j_2}^\prime$. Note that $I_{j_1}^L\cup I_{j_1}^\prime=\sparenv{p_{j_1-1}b-b+2,p_{j_1}b}$ and $I_{j_2}^L\cup I_{j_2}^\prime=\sparenv{p_{j_2-1}b-b+2,p_{j_2}b}$. Then it follows that
\begin{equation}\label{eq_valuecase1}
    x_k=
    \begin{cases}
        x^\prime_k,\text{ if }k<p_{j_1-1}b-b+2,\\
        x^\prime_{k-b},\text{ if }p_{j_1}b<k<p_{j_2-1}b-b+2,\\
        x^\prime_{k-2b},\text{ if }p_{j_2}b<k\le n.
    \end{cases}
\end{equation}
In particular, $\bfx_{I_j^\prime}$ is known to us for any $j\notin\{j_1-1,j_1,j_2-1,j_2\}$. Therefore, to decode $\bfx$ from $\bfx^\prime$, it is sufficient to recover $\bfx_{I_j^\prime}$ for $j\in\{j_1-1,j_1,j_2-1,j_2\}$.

In the rest of this subsection, it is always assumed that $j_2-j_1\ge 2$, since otherwise, we have $\abs{I_{j_1}^L\cup I_{j_1}^\prime\cup I_{j_2}^L\cup I_{j_2}^\prime}\le 2bd\log(n/b)+b-1$,
and therefore, this situation is covered by case (2) in \Cref{lem_qpositions} (or equivalently, case (2) in \Cref{lem_positions}).

Since $j_2-j_1\ge 2$, deletions in $\bfx_{I_{j_2}^L\cup I_{j_2}^\prime}$ do not affect $\bfx_{I_{j_1}^L\cup I_{j_1}^\prime}$. As a result, we have $\bfx^\prime_{\sparenv{p_{j_1-2}b+1,p_{j_1}b-b}}\in\cB_1^b\parenv{\bfx_{I_{j_1-1}^\prime\cup I_{j_1}^\prime}}$ \big(i.e., $\bfx^\prime_{\sparenv{p_{j_1-2}b+1,p_{j_1}b-b}}$ is obtained from $\bfx_{I_{j_1-1}^\prime\cup I_{j_1}^\prime}$ by single $b$-burst-deletion\big) and $\bfx^\prime_{\sparenv{(p_{j_2-2}-1)b+1,p_{j_2}b-2b}}\in\cB_1^b\parenv{\bfx_{I_{j_2-1}^\prime\cup I_{j_2}^\prime}}$ due to the fact that $D_1\subseteq I_{j_1}^L\cup I_{j_1}^\prime\subseteq I_{j_1-1}^\prime\cup I_{j_1}^\prime$ and $D_2\subseteq I_{j_2}^L\cup I_{j_2}^\prime\subseteq I_{j_2-1}^\prime\cup I_{j_2}^\prime$. Applying \Cref{obs_singleburst} to $\bbu=\bfx_{I_{j_1-1}^\prime\cup I_{j_1}^\prime}$ (i.e., the concatenation of $\bfx_{I_{j_1-1}}$ and $\bfx_{I_{j_1}}$), $\bbv=\bfx^\prime_{\sparenv{p_{j_1-2}b+1,p_{j_1}b-b}}$, $m=\parenv{p_{j_1}-p_{j_1-2}}b$ and $i=\parenv{p_{j_1-1}-p_{j_1-2}}b$, we conclude that
\begin{equation}\label{eq_burst1}
    \begin{aligned}
        \bfx^\prime_{\sparenv{p_{j_1-2}b+1,p_{j_1-1}b-b}}\in\cB_1^b\parenv{\bfx_{I_{j_1-1}^\prime}},\\
        \bfx^\prime_{\sparenv{p_{j_1-1}b+1,p_{j_1}b-b}}\in\cB_1^b\parenv{\bfx_{I_{j_1}^\prime}}.
    \end{aligned}
\end{equation}
Similarly, we have 
\begin{equation}\label{eq_burst2}
    \begin{aligned}
        \bfx^\prime_{\sparenv{(p_{j_2-2}-1)b+1,p_{j_2-1}b-2b}}\in\cB_1^b\parenv{\bfx_{I_{j_2-1}^\prime}},\\
        \bfx^\prime_{\sparenv{(p_{j_2-1}-1)b+1,p_{j_2}b-2b}}\in\cB_1^b\parenv{\bfx_{I_{j_2}^\prime}}.
    \end{aligned}
\end{equation}
Recall that the range of $j_1$ is $\sparenv{1,r(U(\bfx)_1)}$. To make (\ref{eq_burst1}) valid when $j_1=1$, we let $\bfx_{I_0^{\prime}}=0^b$ (the sequence consisting of $b$ symbols $0$) and $\bfx^\prime_{\sparenv{p_{-1}b+1,-b}}$ be an empty sequence.

By (\ref{eq_burst1}) and (\ref{eq_burst2}), to recover $\bfx_{I_j^\prime}$ for $j\in\{j_1-1,j_1,j_2-1,j_2\}$, we need to apply to each $\bfx_{I_j^\prime}$ a code for correcting single $b$-burst-deletion.
Let $\phi(\cdot)$ be the function defined in \Cref{lem_singleburst}.  Then by \Cref{lem_singleburst}, our goal boils down to recovering the values of $\phi\parenv{\bfx_{I_{j_1-1}^\prime}}$, $\phi\parenv{\bfx_{I_{j_1}^\prime}}$, $\phi\parenv{\bfx_{I_{j_2-1}^\prime}}$ and $\phi\parenv{\bfx_{I_{j_2}^\prime}}$. Since $U(\bfx)_1$ is $d$-regular and each $U(\bfx)_{1,I_{j}}$ is a run of $U(\bfx)_1$, we have $\abs{I_j}\le d\log(n/b)$ and hence $\abs{I_j^\prime}\le db\log(n/b)$ for all $0\le j\le r(U(\bfx)_1)$. Therefore, each $\phi\parenv{\bfx_{I_j^\prime}}$ can be viewed as an integer in $[0,N-1]$, where $N=2^{\log\log(n)+O_{q,d,b}(1)}$. Here, $O_{q,d,b}(1)$ denotes a constant dependent only on $q,d$ and $b$.

Define $\overline{\phi}(\bfx)=\phi\parenv{\bfx_{I_{0}^\prime}}\phi\parenv{\bfx_{I_{1}^\prime}}\cdots\phi\parenv{\bfx_{I_{r(U(\bfx)_1)}^\prime}}\in[0,N-1]^{r(U(\bfx)_1)+1}$.
According to (\ref{eq_valuecase1}), when given $\bfx^\prime$, the value of $\phi\parenv{\bfx_{I_j^\prime}}$ is known to us for all $j\notin\{j_1-1,j_1,j_2-1,j_2\}$. Therefore, sequence $\overline{\phi}(\bfx)$ suffered two bursts (of length two) of erasures. To correct errors in $\overline{\phi}(\bfx)$, let $\varphi(\cdot)$ be the function defined in \Cref{lem_tberasures} and define
\begin{equation*}
    f(\bfx)=\varphi\parenv{\overline{\phi}(\bfx)}.
\end{equation*}
Recall that we have $r(U(\bfx)_1)\le n/b$. Then By \Cref{lem_tberasures}, $f(\bfx)$ can be viewed as an integer in $[0,N_1-1]$, where $N_1=2^{\log n+4\log\log n+O_{q,d,b}(1)}$.

The following lemma follows from the above analysis, \Cref{lem_singleburst,lem_tberasures}.
\begin{lemma}\label{lem_qruns}
    Suppose that we have found $1\le j_1<j_2\le r\parenv{U(\bfx)_1}$, where $j_2-j_1\ge2$, such that $D_1\subseteq I_{j_1}^L\cup I_{j_1}^\prime$ and $D_2\subseteq I_{j_2}^L\cup I_{j_2}^\prime$. Given $f(\bfx)$, we can efficiently recover $\bfx$ from $\bfx^\prime$.
\end{lemma}

\subsection{When Case (2) in \Cref{lem_qpositions} happens}\label{sec_qinterval}
%%%%%%%%%%%%%%%%%%%%%%%%%%%%%%%%%%%%%%%
Recall that case (2) in \Cref{lem_qpositions} corresponds to case (2) in \Cref{lem_positions}. Therefore, for case (2), instead of looking at $\bfx$ and $\bfx^\prime$, we look at $A(\bfx)$ and $A(\bfx^\prime)$. In this case, we have already decoded $U(\bfx)_1$ from $U(\bfx^\prime)_1$ by using $\eta\parenv{U(\bfx)_1}$. Furthermore, by case (2) in \Cref{lem_positions}, we found an interval $[c_1,c_2]\subseteq[n/b]$ of length at most $3d\log(n/b)$, which contains the positions of the two deletions in $U(\bfx)_1$ (and hence in $A(\bfx)_1$). By \Cref{obs_deletionposition}, the two deletions in $A(\bfx)_i$ occurred in the interval $[c_1-1,c_2]$, for all $2\le i\le b$. Let $A(\bfx)_{i,k}$ be the $k$-th coordinate in the $A(\bfx)_i$. Then $A(\bfx)_{i,k}$ is known to us for all $k\notin [c_1-1,c_2]$. In fact, it holds that $A(\bfx)_{i,k}=A(\bfx^\prime)_{i,k}$ when $1\le k<c_1-1$ and $A(\bfx)_{i,k}=A(\bfx^\prime)_{i,k-2}$ when $c_2<k\le n/b$.

Let $P=3d\log(n/b)+1$. Define
\begin{equation*}
    J_j=
    \begin{cases}
        \sparenv{(j-1)P+1,(j+1)P},\mbox{ if }1\le j\le\ceilenv{n/bP}-2,\\
        \sparenv{(j-1)P+1,n},\mbox{ if }j=\ceilenv{n/bP}-1.
    \end{cases}
\end{equation*}
Since $c_1$ and $c_2$ are known and $\abs{[c_1-1,c_2]}\le 3d\log(n/b)+1$, we can find some $j_0$ such that $[c_1-1,c_2]\subseteq J_{j_0}$. Suppose $J_{j_0}=[d_1,d_2]$. According to above discussion, $A(\bfx)_{i,k}$ is known to us for all $k\notin J_{j_0}$. To decode $A(\bfx)_i$ from $A(\bfx^\prime)_i$, it remains to recover $A(\bfx)_{i,J_{j_0}}$. It is easy to see that $A(\bfx^\prime)_{i,[d_1,d_2-2]}$ is obtained from $A(\bfx)_{i,J_{j_0}}$ by two deletions. Therefore, we can apply \Cref{lem_trivial2deletion} to recover $A(\bfx)_{i,J_{j_0}}$ from $A(\bfx^\prime)_{i,[d_1,d_2-2]}$.

Let $\xi_1(\cdot)$ be the function defined in \Cref{lem_trivial2deletion}.
For each $1\le j\le \ceilenv{n/bP}-1$, let function $g_j(\bfx)$ be defined in \eqref{eq_gj1}.
\begin{figure*}[!t]
    \begin{equation}\label{eq_gj1}
        g_j\parenv{\bfx}=
\begin{cases}
    \parenv{\xi_1\parenv{A(\bfx)_{1,J_j}},\xi_1\parenv{A(\bfx)_{2,J_j}},\xi_1\parenv{A(\bfx)_{3,J_j}},\ldots,\xi_1\parenv{A(\bfx)_{b,J_j}}},\mbox{ if }q>2;\\
    \parenv{\xi_1\parenv{A(\bfx)_{2,J_j}},\xi_1\parenv{A(\bfx)_{3,J_j}},\ldots,\xi_1\parenv{A(\bfx)_{b,J_j}}},\mbox{ if }q=2.
\end{cases}
    \end{equation}\hrulefill
\end{figure*}

Note that in the above definition of $g_j(\bfx)$, we distinguish the two cases where $q>2$ and $q=2$. This is because when $q=2$, we have $A(\bfx)=U(\bfx)$, and hence the first row of $A(\bfx)$ was already recovered.

Since $\abs{J_j}\le 2P=6d\log\parenv{n/b}+2$ for all $j$, by \Cref{lem_trivial2deletion}, each $\xi_1\parenv{A(\bfx)_{i,J_j}}$ is a binary vector of length $7\ceilenv{\log q}\log\log n$ $+O_{q,d,b}(1)$. Therefore, 
each $g_j(\bfx)$ is a binary vector of length at most $7b\ceilenv{\log q}\log\log n+O_{q,d,b}(1)$ (when $q>2$) or $7(b-1)\log\log n+O_{d,b}(1)$ (when $q=2$). Therefore, we can view $g_j(\bfx)$ as an integer in $\sparenv{0,N_2-1}$, where 
\begin{equation}\label{eq_valueN2}
N_2=
\begin{cases}
 2^{7b\ceilenv{\log q}\log\log n+O_{q,d,b}(1)},\mbox{ if }q>2,\\
    2^{7(b-1)\log\log n+O_{d,b}(1)},\mbox{ if }q=2.\\
\end{cases}
\end{equation}

For $s\in\{0,1\}$, let $L_s=\mathset{1\le j\le\ceilenv{n/bP}-1:j\equiv s\pmod{2}}$. For $s\in\{0,1\}$, define
\begin{equation}\label{eq_interval}
    h^{(s)}(\bfx)=\mathop{\sum}\limits_{j\in L_s}g_j(\bfx)\pmod{N_2}.
\end{equation}

\begin{lemma}\label{lem_qinterval}
 For case (2) in \Cref{lem_qpositions} (and hence case (2) in \Cref{lem_positions}), given $h^{(0)}(\bfx)$ and $h^{(1)}(\bfx)$, we can uniquely recover $A(\bfx)_{i,J_{j_0}}$ from $\bfx^\prime$, for all $i$. In particular, we can recover $\bfx$ from $\bfx^\prime$.
\end{lemma}
\begin{IEEEproof}
Without loss of generality, assume $j_0\in L_0$. Then $g_j(\bfx)$ is known to us for each $j\in L_0\setminus\{j_0\}$. By above discussion and \Cref{lem_trivial2deletion}, to recover $A(\bfx)_{i,J_{j_0}}$ for all $2\le i\le b$, it is sufficient to obtain the value of $g_{j_0}(\bfx)$. It follows from \Cref{eq_interval}, that
$$
g_{j_0}(\bfx)\equiv\parenv{h^{(0)}(\bfx)-\mathop{\sum}\limits_{j\in L_0\setminus\{j_0\}}g_{j}(\bfx)}\pmod{N_2}.
$$
Since $0\le g_{j_0}(\bfx)<N_2$, we have
$$
g_{j_0}(\bfx)=\parenv{h^{(0)}(\bfx)-\mathop{\sum}\limits_{j\in L_0\setminus\{j_0\}}g_{j}(\bfx)}\pmod{N_2}.
$$
Now the proof is completed.
\end{IEEEproof}

We are ready to present our main result in this section.
\begin{theorem}\label{thm_qtwoburst}
Suppose $q\ge 2$, $n>2b$ and $b>1$.
Let $N_1$, $N_2$, $f(\bfx)$, $h^{(0)}(\bfx)$ and $h^{(1)}(\bfx)$ be defined as above. Let $N_0=2^{4\log n+10\log\log n+O(1)}$. For any $0\le a<N_0$, $0\le b<N_1$ and $0\le c_0,c_1<N_2$, define the code $\cC_1$ as
\begin{equation*}
    \cC_1\triangleq\mathset{\bfx\in\Sigma_q^n:
    \begin{array}{c}
    A(\bfx)_1\text{ is }d\text{-regular},\\
    \eta(U(\bfx)_1)=a,\\
    f(\bfx)=b,\\
    h^{(0)}(\bfx)=c_0,h^{(1)}(\bfx)=c_1
    \end{array}
    }.
\end{equation*}
Then $\cC_1$ is a $2$-$b$-burst-deletion correcting code. Furthermore, when $(q,n,d)$ satisfies one of the following conditions, there is some $a$, $b$, $c_0$ and $c_1$, such that the redundancy of $\cC_1$ is at most $5\log n+(14b\ceilenv{\log q}+14)\log\log n+O_{q,b}(1)$:
\begin{itemize}
    \item $q>2$ is even, $d=7$ and $n\ge 9$;
    \item $q$ is odd, $d=10$ and $\floorenv{5\log n}n^{-1-5\log(0.87)}\ge \frac{200q}{87(q-1)}$.
\end{itemize}
If $q=2$, $d=7$ and $n\ge 9$, the redundancy is upper bounded by $5\log n+14b\log\log n+O_{b}(1)$.
\end{theorem}
\begin{IEEEproof}
Suppose $\bfx\in\cC_1$ and $\bfx^\prime=\bfx_{[n]\setminus(D_1\cup D_2)}$, where $D_1,D_2\subseteq[n]$ are two disjoint intervals of length $b$. Since $U(\bfx)_1$ is $d$-regular, by \Cref{lem_qpositions}, given $\eta\parenv{U(\bfx)_1}$, we can either find $1\le j_1<j_2\le r\parenv{U(\bfx)_1}$, such that $D_1\subseteq I_{j_1}^L\cup I_{j_1}^\prime$ and $D_2\subseteq \sparenv{p_{j_2-1}b-b+2,p_{j_2-1}b}\cup I_{j_2}^\prime$; or find an interval $J\subseteq[n]$ of length at most $3bd\log(n/b)+b-1$, such that $D_1,D_2\subseteq J$. For the former case, \Cref{lem_qruns} assert that $\bfx$ can be decoded from $\bfx^\prime$. For the latter case, \Cref{lem_qinterval} asserts that $\bfx$ can be decoded from $\bfx^\prime$. Therefore, $\cC_1$ can correct two $b$-burst-deletions.

By \Cref{lem_qregular}, there are at least $q^{n-1}$ sequences $\bfx\in\Sigma_q^n$ such that $A(\bfx)_1$ is $d$-regular. It then follows from the pigeonhole principle that there are some $a$, $b$, $c_0$, and $c_1$, such that
$$
\abs{\cC_1}\ge\frac{q^{n-1}}{N_0\cdot N_1\cdot(N_2)^2},
$$
which implies by definition that
\begin{align*}
    \rho\parenv{\cC_1}&\le\log N_0+\log N_1+2\log N_2+1\\
    &=5\log n+14\log\log n+2\log N_2+O_{q,b}(1).
\end{align*}
Now the claim for redundancy follows from (\ref{eq_valueN2}).
\end{IEEEproof}

\section{Non-binary Two-deletion Correcting Codes}\label{sec_qtwodeletion}
%%%%%%%%%%%%%%%%%%%%%%%%%%%%%%%%%%%%%%%%%%%%
In this section, it is always assumed that $q>2$. Recall that $\cB_t(\bfx)$ is the set of sequences obtained from $\bfx$ by $t$ deletions. We aim to give a new construction of $q$-ary two-deletion correcting codes. Recall the definition of $u(\bfx)$ for any $\bfx\in\Sigma_q^n$. If $\bfx^\prime\in\cB_2(\bfx)$, then $u(\bfx^\prime)\in\cB_2(u(\bfx))$. So if $u(\bfx)$ belongs to a binary two-deletion correcting code, we can recover $u(\bfx)$ from $u(\bfx^\prime)$. We further assume that $\bfx$ (and thus $u(\bfx)$) is $d$-regular. Then  \Cref{lem_positions} asserts that we can either find two runs of $u(\bfx)$ and each of them suffers one deletion, or find a substring of $u(\bfx)$ of short length which suffers two deletions. This enables us to approximately determine error positions in $\bfx$. For the latter case, we need a $q$-ary two-deletion correcting code, which is given in \Cref{lem_trivial2deletion}. For the former case, we need a $q$-ary single-deletion correcting code. We use the code given in \cite{Tuan2024IT}.
\begin{lemma}\cite{Tuan2024IT}\label{lem_DVT}
    Suppose $q>2$ and $n>2$. There is a function $\dvt:\Sigma_q^n\rightarrow\Sigma_2^{\log n+\log q}$, computable in linear time, such that for any $\bfx\in\Sigma_q^n$, given $\dvt(\bfx)$ and $\bby\in\cB_1(\bfx)$, one can uniquely and efficiently recover $\bfx$.
\end{lemma}

Let $u(\bfx)_{I_j}$ ($1\le j\le r\parenv{u(\bfx)}$) be all the runs in $u(\bfx)$. Furthermore, suppose $I_j=\sparenv{p_{j-1}+1,p_{j}}$, where $p_0=0$, $p_{r\parenv{u(\bfx)}}=n$ and $p_{j-1}<p_j$ for all $j$. Since $u(\bfx)$ is $d$-regular, we have $\abs{I_j}\le d\log n$. Therefore, by \Cref{lem_DVT}, $\dvt\parenv{\bfx_{I_j}}$ can be viewed as an integer in $\sparenv{0,N_1-1}$, where $N_1=2^{\log(d\log n)+\log q}$. Let $Q\ge\max\{n,N_1\}$ be the smallest prime. By the following lemma, we have $n\le Q<2n$.
\begin{lemma}[Bertrand–Chebyshev theorem]\label{lem_primes}
    For every integer $n\ge2$, there is always at least one prime $p$ such that $n\le p<2n$.
\end{lemma}
Define
\begin{equation}\label{eq_delrun}
    \begin{aligned}
        f_0(\bfx)=\sum_{j=1}^{r\parenv{u(\bfx)}}\dvt\parenv{\bfx_{I_j}}\pmod{2N_1},\\
        f_1(\bfx)=\sum_{j=1}^{r\parenv{u(\bfx)}}j\dvt\parenv{\bfx_{I_j}}\pmod{Q}.
    \end{aligned}
\end{equation}
The two functions $f_0(\bfx)$ and $f_1(\bfx)$ can help to deal with case (1) in \Cref{lem_positions}, as shown in the next lemma.
\begin{lemma}\label{lem_2deletion1}
     Suppose that the two deletions in $\bfx$ occurred in two known intervals $I_{j_1}=\sparenv{p_{j_1-1}+1,p_{j_1}}$ and $I_{j_2}=\sparenv{p_{j_2-1}+1,p_{j_2}}$, respectively. Then given $f_0(\bfx)$ and $f_1(\bfx)$, we can uniquely recover $\bfx$ from $\bfx^\prime$.
\end{lemma}
\begin{IEEEproof}
   Notice that $x_k$ is known to us for all $k\notin I_{j_1}\cup I_{j_2}$. It remains to recover $\bfx_{I_{j_1}}$ and $\bfx_{I_{j_2}}$. It is easy to see that $\bfx^\prime_{\sparenv{p_{j_1-1}+1,p_{j_1}-1}}\in\cB_1\parenv{\bfx_{I_{j_1}}}$ and $\bfx^\prime_{\sparenv{p_{j_1-2},p_{j_2}-2}}\in\cB_1\parenv{\bfx_{I_{j_2}}}$. By \Cref{lem_DVT}, to recover $\bfx_{I_{j_1}}$ and $\bfx_{I_{j_2}}$, it is sufficient to know the values of $\dvt\parenv{\bfx_{I_{j_1}}}$ and $\dvt\parenv{\bfx_{I_{j_2}}}$. Note that $\dvt\parenv{\bfx_{I_{j}}}$ is known to us for all $j\ne j_1,j_2$. Let $\delta_0=\parenv{f_0(\bfx)-\sum_{j\ne j_1,j_2}\dvt\parenv{\bfx_{I_j}}}\pmod{2N_1}$ and $\delta_1=\parenv{f_1(\bfx)-\sum_{j\ne j_1,j_2}j\dvt\parenv{\bfx_{I_j}}}\pmod{Q}$. Then it follows from \Cref{eq_delrun} that
\begin{align}           
 \dvt\parenv{\bfx_{I_{j_1}}}+\dvt\parenv{\bfx_{I_{j_2}}}\equiv\delta_0\pmod{2N_1},\label{eq_run1}\\
 j_1\dvt\parenv{\bfx_{I_{j_1}}}+j_2\dvt\parenv{\bfx_{I_{j_2}}}\equiv\delta_1\pmod{Q}\label{eq_run2}.
\end{align}
Since $0\le \dvt\parenv{\bfx_{I_{j_1}}}+\dvt\parenv{\bfx_{I_{j_2}}}<2N_1$ and $0\le\delta_0<2N_1$, we can conclude from \Cref{eq_run1} that $\dvt\parenv{\bfx_{I_{j_1}}}+\dvt\parenv{\bfx_{I_{j_2}}}=\delta_0$. Combining this with \Cref{eq_run2}, we get
\begin{equation}\label{eq_run3}
    \begin{aligned}
        \dvt\parenv{\bfx_{I_{j_1}}}+\dvt\parenv{\bfx_{I_{j_2}}}\equiv\delta_0\pmod{Q},\\
 j_1\dvt\parenv{\bfx_{I_{j_1}}}+j_2\dvt\parenv{\bfx_{I_{j_2}}}\equiv\delta_1\pmod{Q}.
    \end{aligned}
\end{equation}
Since $1\le j_1<j_2\le n\le Q$, we have $j_1\not\equiv j_2\pmod{Q}$. Therefore, system (\ref{eq_run3}) has a unique solution in the field $\mathbb{F}_Q$: $\parenv{\dvt\parenv{\bfx_{I_{j_1}}}\pmod{Q},\dvt\parenv{\bfx_{I_{j_2}}}\pmod{Q}}$. Since $0\le\dvt\parenv{\bfx_{I_{j_1}}},\dvt\parenv{\bfx_{I_{j_2}}}<N_1\le Q$, we have $\dvt\parenv{\bfx_{I_{j_1}}}=\dvt\parenv{\bfx_{I_{j_1}}}\pmod{Q}$ and $\dvt\parenv{\bfx_{I_{j_2}}}=\dvt\parenv{\bfx_{I_{j_2}}}\pmod{Q}$. Now the proof is completed.
\end{IEEEproof}

\bigskip
For case (2) in \Cref{lem_positions}, we follow similar idea in \Cref{sec_qinterval}. Let $P=3d\log n$ and
\begin{equation*}
    J_j=
    \begin{cases}
        \sparenv{(j-1)P+1,(j+1)P},\mbox{ if }1\le j\le\ceilenv{n/P}-2,\\
        \sparenv{(j-1)P+1,n},\mbox{ if }j=\ceilenv{n/P}-1.
    \end{cases}
\end{equation*}
Let $\xi_1(\cdot)$ be the function in \Cref{lem_trivial2deletion}, which can help to correct two deletions in $q$-ary sequences. Since $\abs{J_1}\le 2P=6d\log n$, we can view each $\xi_1\parenv{\bfx_{J_j}}$ as an integer in $\sparenv{0,N_2-1}$, where $N_2=2^{7\ceilenv{\log q}\log(6d\log n)+O_q(1)}$ (see \Cref{lem_trivial2deletion}). For $s\in\{0,1\}$, let $K_s=\mathset{1\le j\le\ceilenv{n/P}-1:j\equiv s\pmod{2}}$ and define
\begin{equation}\label{eq_delinterval}
    h^{(s)}(\bfx)=\mathop{\sum}\limits_{j\in K_s}\xi_1\parenv{\bfx_{J_j}}\pmod{N_2}.\\
\end{equation}
\begin{lemma}\label{lem_2deletion2}
    Suppose we are in case (2) in \Cref{lem_positions}. Given $h^{(0)}$ and $h^{(1)}$, we can uniquely recover $\bfx$ from $\bfx^\prime$.
\end{lemma}
The proof is similar to that of \Cref{lem_qinterval}, and thus omitted.

Our $q$-ary two-deletion correcting code is given in the next theorem. 
\begin{theorem}\label{thm_qtwodeletion}
Suppose $q> 2$.
Let $N_1,N_2,Q$, $f_0(\bfx)$, $f_1(\bfx)$, $h^{(0)}(\bfx)$ and $h^{(1)}(\bfx)$ be defined as above. For all $0\le a<2^{4\log n+10\log\log n+O(1)}$, $0\le b_{0}<2N_1$, 
$0\le b_{1}<Q$, and $0\le c_0,c_1<N_2$, define the code $\cC_2$ as
\begin{equation*}
    \cC_2\triangleq\mathset{\bfx\in\Sigma_q^n:
    \begin{array}{c}
    \bfx\text{ is }d\text{-regular},\\
    \eta(u(\bfx))=a,\\
    f_s(\bfx)=b_{s}\text{ for
    }s\in\{0,1\},\\
    h^{(0)}(\bfx)=c_0,h^{(1)}(\bfx)=c_1
    \end{array}
    }.
\end{equation*}
Then $\cC_2$ is a two-deletion correcting code. Furthermore, when $(q,n,d)$ satisfies one of the following conditions, there are some $a$, $b_{0}$, $b_1$, $c_0$ and $c_1$, such that the redundancy of $\cC_2$ is at most $5\log n+(14\ceilenv{\log q}+11)\log\log n+O_{q}(1)$:
\begin{itemize}
    \item $q$ is even, $d=7$ and $n\ge 9$;
    \item $q$ is odd, $d=10$ and $\floorenv{5\log n}n^{-1-5\log(0.87)}\ge \frac{200q}{87(q-1)}$.
\end{itemize}
\end{theorem}
The proof of \Cref{thm_qtwodeletion} follows directly from \Cref{lem_2deletion1,lem_2deletion2}. The claim for redundancy follows from \Cref{lem_qregular} and \Cref{lem_primes}.

\section{Conclusion}\label{sec_conclusion}
%%%%%%%%%%%%%%%%%%%%%%%%%%%%%%%%%%%%%%%%%%%%%%%
In this paper, we present constructions of $q$-ary codes capable of correcting two bursts of exactly $b$ deletions with redundancy at most $5\log n+O(\log\log n)$ for all $b\ge 2$ and $q\ge 2$, which improves the redundancy of codes derived from the syndrome compression technique. Inspired by these ideas, we provide a new construction of $q$-ary two-deletion correcting codes with redundancy $5\log n+O(\log\log n)$ for all $q>2$.

In this work, it is required that the two bursts have the same length. Allowing the two deletion-bursts to have different sizes would make the error model more general and applicable to a wider range of practical scenarios. Indeed, such a generalization would be more realistic, as burst deletions in real-world applications (e.g., communication systems or storage systems) may not always have equal lengths. The case where the two bursts have different lengths seems more involved. Extending our work to handle bursts of different sizes is a promising direction for future research.

A more complex problem is to construct codes capable of correcting at most two bursts of deletions, where each burst has a length at most $b$. Our technique fails in these setup since we can not even know the number of bursts or the length of each burst occurring in a received sequence. This problem is also left for future research. 

\section*{Acknowledgement}
The authors express their gratitude to the anonymous reviewers for their detailed and constructive comments which are very helpful to the improvement of the presentation of this paper. The authors would also like to thank Prof. Alberto Ravagnani, the associate editor, for his excellent editorial job.

\appendices
\section{A Lemma}\label{appendix1}
    The following \Cref{lem_A1} is derived by slightly modifying the proof of \cite[Proposition 3.3.7]{Guruswami2023}. Before showing its proof, we list some facts.
    Let $\ln(\cdot)$ be the natural logarithm. Then when $\abs{x}<1$, we have $\ln(1+x)=\sum_{i=1}^{\infty}(-1)^{i-1}\frac{x^i}{i}=x-\frac{x^2}{2}+\frac{x^3}{3}-\cdots$.
\begin{fact}\label{fact_appendix1}
For all $0\le x<1$, we have $\ln(1-x)\le -x-\frac{x^2}{2}$ and $-\ln(1+x)\le -x+\frac{x^2}{2}$.
When $0\le x\le\frac{1}{2}$, we have $-\ln(1-x)\le x+\frac{x^2}{2}+2x^3$.
\end{fact}
\begin{IEEEproof}
    When $0\le x<1$, we have $\ln(1-x)=\sum_{i=1}^{\infty}(-1)^{i-1}\frac{(-x)^i}{i}=-\sum_{i=1}^{\infty}\frac{x^i}{i}\le -x-\frac{x^2}{2}$. Since $-\ln(1+x)=-x+\frac{x^2}{2}-\parenv{\ln(1+x)-x+\frac{x^2}{2}}$ and $\ln(1+x)-x+\frac{x^2}{2}\ge0$ for all $x\ge 0$, we have $-\ln(1+x)\le -x+\frac{x^2}{2}$. At last, it is easy to see that $-\ln(1-x)=x+\frac{x^2}{2}+x^3\parenv{\frac{1}{3}+\frac{x}{4}+\frac{x^2}{5}+\cdots}\le x+\frac{x^2}{2}+x^3(1+x+x^2+\cdots)=x+\frac{x^2}{2}+\frac{x^3}{1-x}$. When $x\le\frac{1}{2}$, we have $1-x\ge\frac{1}{2}$ and thus $-\ln(1-x)\le x+\frac{x^2}{2}+2x^3$.
\end{IEEEproof}

\begin{lemma}\label{lem_A1}
Suppose that $q\ge2$ is fixed and $\epsilon<\min\mathset{\frac{1}{q},\frac{q-1}{2q},\frac{(q-1)^2}{q^2-3q+6}\parenv{\frac{1}{q}-\frac{(q-1)\ln q}{2q^3}}}$. Then it holds that
$$
H_q(1 - \frac{1}{q} - \epsilon)\leq 1 - \frac{\epsilon^2}{4}.
$$
\end{lemma}
\begin{IEEEproof}
It is easy to verify that
    \begin{equation}\label{eq_appendix1}
        \begin{aligned}
            & H_q\parenv{1-\frac{1}{q}-\epsilon}\\
            =&-\parenv{1-\frac{1}{q}-\epsilon}\log_q\parenv{\frac{1-1/q-\epsilon}{q-1}}\\
            &\quad-\parenv{\frac{1}{q}+\epsilon}\log_q\parenv{\frac{1}{q}+\epsilon}\\
            =&-\log_q\parenv{\frac{1}{q}\parenv{1-\frac{\epsilon q}{q-1}}}\\
            &\quad+\parenv{\frac{1}{q}+\epsilon}\log_q\parenv{\frac{1-(\epsilon q)/(q-1)}{1+\epsilon q}}\\
            =&1-\frac{1}{\ln q}\left[\ln\parenv{1-\frac{\epsilon q}{q-1}}\right.\\
            &\quad\left.-\parenv{\frac{1}{q}+\epsilon}\ln\parenv{\frac{1-(\epsilon q)/(q-1)}{1+\epsilon q}}\right]\\
            =&1+\frac{1}{\ln q}\left[-\ln\parenv{1-\frac{\epsilon q}{q-1}}+\parenv{\frac{1}{q}+\epsilon}\ln\parenv{1-\frac{\epsilon q}{q-1}}\right.\\
            &\quad\quad\quad\quad\quad\left.-\parenv{\frac{1}{q}+\epsilon}\ln\parenv{1+\epsilon q}\right].
        \end{aligned}
\end{equation}
Since $\epsilon<\min\mathset{\frac{1}{q},\frac{q-1}{2q}}$, we have $0<\frac{\epsilon q}{q-1}<\frac{1}{2}$ and $0<\epsilon q<1$. Then it follows from \Cref{fact_appendix1} that $-\ln\parenv{1-\frac{\epsilon q}{q-1}}\le\frac{\epsilon q}{q-1}+\frac{\epsilon^2 q^2}{2(q-1)^2}+\frac{2\epsilon^3 q^3}{(q-1)^3}$, $\ln\parenv{1-\frac{\epsilon q}{q-1}}\le-\frac{\epsilon q}{q-1}-\frac{\epsilon^2 q^2}{2(q-1)^2}$ and $-\ln(1+\epsilon q)\le-\epsilon q+\frac{\epsilon^2 q^2}{2}$. Now by \Cref{eq_appendix1}, we have
\begin{equation*}
\begin{aligned}
           &H_q\parenv{1-\frac{1}{q}-\epsilon}\\
           \le&1+\frac{2\epsilon^3 q^3}{(q-1)^3\ln q}+\frac{1}{\ln q}\left[\frac{\epsilon q}{q-1}+\frac{\epsilon^2 q^2}{2(q-1)^2}\right.\\
           &\quad\quad\left.+\parenv{\frac{1}{q}+\epsilon}\parenv{-\frac{\epsilon q}{q-1}-\frac{\epsilon^2 q^2}{2(q-1)^2}-\epsilon 
 q+\frac{\epsilon^2 q^2}{2}}\right]\\
 =&1+\frac{2\epsilon^3 q^3}{(q-1)^3\ln q}+\frac{1}{\ln q}\left[\frac{\epsilon q}{q-1}+\frac{\epsilon^2 q^2}{2(q-1)^2}\right.\\
 &\quad\quad\quad\quad\quad\quad\quad\left.+\parenv{\frac{1}{q}+\epsilon}\parenv{-\frac{\epsilon q^2}{q-1}+\frac{\epsilon^2 q^3(q-2)}{2(q-1)^2}}\right]\\
 =&1+\frac{\epsilon^3 q^3(q^2-3q+6)}{2(q-1)^3\ln q}\\
 &\quad\quad\quad\quad\quad+\frac{1}{\ln q}\sparenv{\frac{\epsilon^2 q^2}{2(q-1)^2}-\frac{\epsilon^2 q^2}{q-1}+\frac{\epsilon^2 q^2(q-2)}{2(q-1)^2}}\\
 =&1-\frac{\epsilon^2 q^2}{2(q-1)\ln q}+\frac{\epsilon^3 q^3(q^2-3q+6)}{2(q-1)^3\ln q}\\
 \overset{(a)}{\le}&1-\frac{\epsilon^2}{4},
        \end{aligned}
\end{equation*}
where (a) follows from the fact that $\epsilon<\frac{(q-1)^2}{q^2-3q+6}\parenv{\frac{1}{q}-\frac{(q-1)\ln q}{2q^3}}$.
\end{IEEEproof}

\section{Encoding sequences into regular sequences}\label{appendix2}
%%%%%%%%%%%%%%%%%%%%%%%%%%%%%%
In this section, we give two methods to encode a sequence into a $d$-regular sequence using only one redundant symbol. The first method is an instantiation of the encoding idea in the proof of \cite[Lemma 11]{Guruswami2021it}. The second method relies on the sequence replacement technique. The first method works for more flexible parameters $n$ and $d$. The time complexity of the second method is lower.

For any $q\ge 2$, let $Q\subseteq\Sigma_q^m$ be the set of sequences containing two consecutive coordinates smaller than $\ceilenv{q/2}$ and two consecutive coordinates no less than $\ceilenv{q/2}$. According to \cite[Lemma 10]{Guruswami2021it}, it holds that $\abs{Q}\ge 2^m-2\times(1.62)^{m+2}$ when $q=2$. By the definition of $q$-ary $d$-regular sequences, it is clear that $\abs{Q}\ge q^m-2\times(1.62)^2\times(0.81q)^m$ for any even $q$. In \Cref{lem_qregular}, it is shown that $\abs{Q}\ge q^m-2\times(0.87q)^m$. 

 \subsection{The first method}
%%%%%%%%%%%%%%%%%%%%%%%%%%%%%%%%%%%
Suppose that $d$ and $n$ satisfy conditions in \Cref{lem_qregular}. In this subsection, let $m=\floorenv{\frac{d}{2}\log n}$, $k=\floorenv{n/m}$ and
\begin{equation*}
\begin{aligned}
 \cR^\prime=\bigg\{\bfx=\bfx^{(0)}\cdots\bfx^{(k-1)}&\bfx^{(k)}\in\Sigma_q^n:\\
&\begin{array}{c}
    \bfx^{(i)}\in Q,\forall 0\le i\le k-1  \\
     \bfx^{(k)}\in\Sigma_q^{n-km}
 \end{array}
 \bigg\}.
 \end{aligned}
 \end{equation*}
 Then in the proof of \cite[Lemma 11]{Guruswami2021it} and \Cref{lem_qregular}, it is shown that each sequence in $d$-regular and $\abs{\cR^\prime}\ge q^{n-1}$.
We aim to encode a sequence of length $n-1$ into $\cR^\prime$. Before showing the algorithm, we need to define some functions.

For $\bfx\in\Sigma_{q}^{n-1}$, define $f_{\textup{dec}}(\bfx)=\sum_{i=1}^{n-1}x_iq^{i-1}$. It is clear that $f_{\textup{dec}}(\bfx)\in\sparenv{0,q^{n-1}-1}$ and $f_{\textup{dec}}(\bfx)$ can be computed in $O(n)$ time.

We further assume that $n$ is sufficiently large such that $m\ge 8$ when $q$ is odd, or such that $m\ge 12$ when $q$ is even. Under this assumption, it is easy to verify that $\abs{Q}\ge q^{m-1}\ge q^{n-km}$.
Since $\abs{Q}^k-1+\abs{Q}^k\parenv{q^{n-km}-1}= \abs{\cR^\prime}-1\ge q^{n-1}-1$, each integer $a$ in $\sparenv{0,q^{n-1}-1}$ can be uniquely written as $a=\sum_{i=0}^{k}a_i\abs{Q}^i$, where $0\le a_0, a_1,\ldots,a_{k-1}<\abs{Q}$ and $0\le a_k<q^{n-km}\le \abs{Q}$. Denote $g_{Q}(a)=\parenv{a_0,\ldots,a_{k-1},a_k}$. The time complexity of computing $g_{Q}(a)$ is $O\parenv{\log_{\abs{Q}}\parenv{q^{n-1}-1}}=O\parenv{\log\parenv{q^{n-1}-1}/\log\abs{Q}}=O\parenv{n/m}$.

Since $\abs{Q}\le q^m=O\parenv{n^{d\log q/2}}$, we can efficiently construct the set $Q$ by brute-force. There is a bijection between the set $Q$ and the integer set $\sparenv{0,\abs{Q}-1}$. We can build a lookup table which gives such a bijection. Then for each sequence in $Q$ we can find its corresponding integer in $\sparenv{0,\abs{Q}-1}$ in $O\parenv{n^{d\log q/2}}$ time. Similarly, there is a bijection between $\Sigma_q^{n-km}$ and the set $\sparenv{0,q^{n-km}-1}$ and for each sequence in $\Sigma_q^{n-km}$, we can find its corresponding integer in $\sparenv{0,q^{n-km}-1}$ in $O\parenv{n^{d\log q/2}}$ time by a lookup table. 

Our encoding algorithm is described as follows. For a sequence $\bfx\in\Sigma_q^{n-1}$, compute $f_{\textup{dec}}(\bfx)$ and $g_{Q}\parenv{f_{\textup{dec}}(\bfx)}$. Suppose that $g_{Q}\parenv{f_{\textup{dec}}(\bfx)}=\parenv{a_0,\ldots,a_{k-1},a_k}$, where $0\le a_0,\ldots,a_{k-1}<\abs{Q}$ and $0\le a_k<q^{n-km}$. As analyzed above, we can map $a_i$ to a sequence $\bby^{(i)}$ in $Q$ for each $0\le i\le k-1$, and map $a_k$ to a sequence $\bby^{(k+1)}$ in $\Sigma_q^{n-km}$. At last, concatenate these sequences and we get the output $\bby=\bby^{(0)}\cdots\bby^{(k-1)}\bby^{(k)}$, which is $d$-regular. Clearly, the time complexity of the encoding process is $O\parenv{(k+1)n^{d\log q/2}}=O\parenv{n^{d\log q/2+1}/m}$.

Decoding $\bby$ to the original sequence is straightforward. And the time complexity is also $O\parenv{n^{d\log q/2+1}/m}$.

\subsection{The second method}
%%%%%%%%%%%%%%%%%%%%%%%%%%%%%%%%%%%%
In this subsection, we give a new encoding algorithm based on the sequence replacement technique. This technique has been widely used in the literature to encode a sequence to a sequence without forbidden substrings \cite{Schoeny2017it,Ohad2021IT,Shuche2024IT,Zuo2024IT}.

Let $m=\floorenv{\frac{d}{3}\log n}$ (where $d$ will be specified later) and $Q$ be the set defined at the beginning of this section. Then we have that $\abs{\Sigma_q^m\setminus Q}\le (c\cdot q)^m$ for some $0<c<1$. Therefore, there is an injective mapping $f_{Q}$ from $\Sigma_q^{m}\setminus Q$ to the set $\Sigma_q^{\ceilenv{m\parenv{1+\log_q c}}}$. Then the inverse mapping $f_{Q}^{-1}: \mathset{f_{Q}:\bbz\in \Sigma_q^{m}\setminus Q}\rightarrow \Sigma_q^{m}\setminus Q$ is well-defined. In addition, by building a lookup table, both $f_{Q}$ and $f_{Q}^{-1}$ can be computed in $O\parenv{(cq)^m}=O\parenv{n^{\frac{d}{3}\log(cq)}}$ time.

In the rest of this subsection, denote $k=\floorenv{n/m}$. Fix a suitable $d$ such that when $n$ is sufficiently large, we have $m\ge \ceilenv{m\parenv{1+\log_q c}}+\ceilenv{\log_q k}+4$.

For a sequence $\bfx\in\Sigma_q^{n-1}$, we first append a symbol $0$ to $\bfx$ and get the sequence $\tilde{\bfx}=\bfx0$. This appended symbol $0$ will help to indicate when our decoding algorithm ends. Next, we partition sequence $\tilde{\bfx}$ as $\tilde{\bfx}=\tilde{\bfx}^{(1)}\cdots\tilde{\bfx}^{(k)}\tilde{\bfx}^{(k+1)}$, where $\tilde{\bfx}^{(i)}\in\Sigma_q^m$ for each $1\le i\le k$ and $\tilde{\bfx}^{(k+1)}\in\Sigma_q^{n-km}$. For any symbol $\alpha\in\Sigma_q$ and integer $\ell\ge 1$, let $\alpha^{\ell}$ denote the sequence consisting of $\ell$ symbols $\alpha$. For simpler notations, denote $m_{1}=\floorenv{\frac{m-\ceilenv{m\parenv{1+\log_q c}}-\ceilenv{\log_q k}}{2}}$ and $m_2=\ceilenv{\frac{m-\ceilenv{m\parenv{1+\log_q c}}-\ceilenv{\log_q k}}{2}}$.

The idea behind the encoding algorithm is as follows.
\begin{enumerate}[\textbf{Step} 1]
    \item For each $1\le i\le k$, check if $\tilde{\bfx}^{(i)}\in Q$. If not, delete $\tilde{\bfx}^{(i)}$ from $\tilde{\bfx}$ and still denote the resulted sequence by $\tilde{\bfx}$.
    \item Then we append the sequence $f_{Q}\parenv{\tilde{\bfx}^{(i)}}b_q(i)0^{m_1}(q-1)^{m_2}$ to the end of $\tilde{\bfx}$. Here, we use $b_q(i)$ to denote the $q$-ary representation of integer $i$. Clearly, $b_q(i)\in\Sigma_q^{\ceilenv{\log_q k}}$.
    \item Note that the substring $\tilde{\bfx}^{(k+1)}$ will never be deleted after Step 1 and Step 2. Continue Step 1 and Step 2 until the left of $\tilde{\bfx}^{(k+1)}$ is empty or all $\tilde{\bfx}^{(i)}$ to the left of $\tilde{\bfx}^{(k+1)}$ is in $Q$.
\end{enumerate}

Since $m\ge \ceilenv{m\parenv{1+\log_q c}}+\ceilenv{\log_q k}+4$, we have $m_2\ge m_1\ge 2$. Therefore, once Step 2 is executed, a sequence in $Q$ is appended to the end. Note that the length of $f_{Q}\parenv{\tilde{\bfx}^{(i)}}b_q(i)0^{m_1}(q-1)^{m_2}$ is $m$. This implies that the length of $\tilde{\bfx}$ does not change after Step 2.
Denote the sequence at the end of this algorithm by $\bby$. The fact that $m=\floorenv{\frac{d}{3}\log n}$ implies $d\log n>3m-3$. Then it is easy to see that any substring of $\bby$ of length at least $d\log n$ either contains a $\tilde{\bfx}^{(i)}$ (which is on the left of $\tilde{\bfx}^{(k+1)}$) or contains a substring of the form $f_{Q}\parenv{\tilde{\bfx}^{(i)}}b_q(i)0^{m_1}(q-1)^{m_2}$ (which is on the right of $\tilde{\bfx}^{(k+1)}$). Therefore, $\bby$ is $d$-regular.

The formal algorithm is given in \Cref{alg_dregular}. Note that in \Cref{alg_dregular}, it always holds that $\tilde{\bfx}^{(j+1)}=\tilde{\bfx}^{(k+1)}$. In each loop, if $\tilde{\bfx}^{(i)}\in\Sigma_q^m\setminus Q$, a block of length $m$ to the left of $\tilde{\bfx}^{(j+1)}$ is deleted. Therefore, in Line 6, the substring between $\tilde{\bfx}^{(j+1)}$ and $f_{Q}\parenv{\tilde{\bfx}^{(i)}}b_q(i)0^{m_1}(q-1)^{m_2}$ is $\tilde{\bfx}_{\sparenv{n-(k-j)m+1,n}}$, which the suffix of $\tilde{\bfx}$ (in previous loop) of length $(k-j)m$. There are at most $k$ loops. In each loop, if $\tilde{\bfx}^{(i)}\in\Sigma_q^m\setminus Q$, we have to compute $f_{Q}\parenv{\tilde{\bfx}^{(i)}}$ and $b_q(i)$. Therefore, the overall time complexity is $O\parenv{kn^{\frac{d}{3}\log(cq)}}=O\parenv{n^{\frac{d}{3}\log(cq)+1}/m}$.

Denote the encoder based on \Cref{alg_dregular} by $\textup{Enc}(\cdot)$. We present the decoding algorithm in \Cref{alg_decoder}, where the function $\textup{Dec}(\cdot)$ computes the decimal representation of an input sequence.
In the Initialization step of \Cref{alg_dregular}, a symbol $0$ is appended to $\bfx$. In Line 6 of \Cref{alg_dregular}, the appended block ends with symbol $q-1$. Therefore, in \Cref{alg_decoder}, the fact that $x_n=q-1$ implies that there is an appended block which has not been deleted and the decoder has to go into another while loop. 
Suppose that
$
\bfx=\tilde{\bfx}^{(1)}\cdots\tilde{\bfx}^{(i-1)}\tilde{\bfx}^{(i+1)}\cdots$ $\tilde{\bfx}^{(j)}\tilde{\bfx}^{(j+1)}\tilde{\bfx}_{\sparenv{n-(k-j)m+1,n}} f_{Q}\parenv{\tilde{\bfx}^{(i)}}b_q(i)0^{m_1}(q-1)^{m_2}.
$
Then it is clear that $\bfx_{\sparenv{n-m_1-m_2-\ceilenv{\log_q k}+1,n-m_1-m_2}}=b_q(i)$ and $\bfx_{\sparenv{n-m_1-m_2-\ceilenv{\log_q k}-\ceilenv{m\parenv{1+\log_q c}}+1,n-m_1-m_2-\ceilenv{\log_q k}}}$. Therefore, we get the index $i$ in Line 4 and $f_{Q}^{-1}$ is the deleted $\tilde{\bfx}^{(i)}$. When the while loop ends, we get the sequence $\bfx0$. In Line 8, the last symbol $0$ is deleted. Now the correctness of \Cref{alg_decoder} is clear. There are at most $k$ while loops and in each loop, we have to compute $\textup{Dec}(\cdot)$ and $f_Q^{-1}(\bbu)$. Therefore, the overall time complexity is $O\parenv{kn^{\frac{d}{3}\log(cq)}}=O\parenv{n^{\frac{d}{3}\log(cq)+1}/m}$.
\begin{algorithm*}[!htbp]
\small{
\DontPrintSemicolon
\SetAlgoLined
\KwIn {$\bfx\in\Sigma_q^{n-1}$}
\KwOut {$\tilde{\bfx}\in\cR_{q,n,d}$}
\textbf{Initialization}\;
$\tilde{\bfx}\gets\bfx0$, write $\tilde{\bfx}=\tilde{\bfx}^{(1)}\cdots\tilde{\bfx}^{(k)}\tilde{\bfx}^{(k+1)}$, where $\tilde{\bfx}^{(i)}\in\Sigma_q^m$ for each $1\le i\le k$ and $\tilde{\bfx}^{(k+1)}\in\Sigma_q^{n-km}$\;
$i\gets 1$, $j\gets k$\;
\While{$i\le j$}
{
\eIf{$\tilde{\bfx}^{(i)}\in\Sigma_q^m\setminus Q$}{
$\tilde{\bfx}\gets\tilde{\bfx}^{(1)}\cdots\tilde{\bfx}^{(i-1)}\tilde{\bfx}^{(i+1)}\cdots\tilde{\bfx}^{(j)}\tilde{\bfx}^{(j+1)}\tilde{\bfx}_{\sparenv{n-(k-j)m+1,n}} f_{Q}\parenv{\tilde{\bfx}^{(i)}}b_q(i)0^{m_1}(q-1)^{m_2}$\;
$j\gets j-1$
}{
$i\gets i+1$
}
}
\Return{$\tilde{\bfx}$}
\caption{Encoding a sequence into a $d$-regular sequence}
\label{alg_dregular}}
\end{algorithm*}

\begin{algorithm}[!htbp]
\small{
\DontPrintSemicolon
\SetAlgoLined
\KwIn {$\tilde{\bfx}=\textup{Enc}(\bfx)\in\Sigma_q^{n-1}$}
\KwOut {$\bfx\in\Sigma_{q}^{n-1}$}
\textbf{Initialization}\;
$\bfx\gets\tilde{\bfx}$\;
\While{$x_n=q-1$}
{
$i_0\gets\textup{Dec}\parenv{\bfx_{\sparenv{n-m_1-m_2-\ceilenv{\log_q k}+1,n-m_1-m_2}}}$\;
$\bbu\gets\bfx_{\sparenv{n-m_1-m_2-\ceilenv{\log_q k}-\ceilenv{m\parenv{1+\log_q c}}+1,n-m_1-m_2-\ceilenv{\log_q k}}}$\;
$\bfx\gets\bfx_{\sparenv{1,(i_0-1)m}}f_{Q}^{-1}\parenv{\bbu}\bfx_{\sparenv{(i_0-1)m+1,n-m}}$
}
$\bfx\gets\bfx_{[1,n-1]}$\;
\Return{$\bfx$}
\caption{Decoding a $d$-regular sequence into the original sequence}
\label{alg_decoder}}
\end{algorithm}

\bibliographystyle{IEEEtran}
\bibliography{ref}

\begin{IEEEbiographynophoto}{Zuo~Ye}
(\emph{Member, IEEE}) received his Ph.D. degree in mathematics from the University of Science and Technology of China, Hefei,
Anhui, P. R. China, in 2021. From 2022 to 2025, he held a postdoctoral position with the School of Electrical and Computer Engineering, Ben-Gurion University of the Negev. He is currently an Associate Professor with the Institute of Mathematics and Interdisciplinary Sciences, Xidian University,
Xi'an 710126, China. His research interests include
combinatorics, coding theory, and their interactions.
\end{IEEEbiographynophoto}

\begin{IEEEbiographynophoto}{Yubo~Sun}
is currently pursuing the Ph.D. degree with Capital Normal
University, Beijing, China. His research interests include combinatorics and coding theory and their interactions.
\end{IEEEbiographynophoto}

\begin{IEEEbiographynophoto}{Wenjun~Yu}
received his Ph.D. degree in mathematics from the University of Science and Technology of China, Hefei,
Anhui, P. R. China, in 2021. Currently, he is holding a postdoctoral position with the School of Electrical and Computer Engineering, Ben-Gurion University of the Negev. His research interests include
combinatorics, graph theory, coding theory, and
their interactions.
\end{IEEEbiographynophoto}

\begin{IEEEbiographynophoto}{Gennian~Ge}
received the M.S. and Ph.D. degrees in mathematics from Suzhou University, Suzhou, Jiangsu, China, in 1993 and 1996, respectively.

After that, he became a member of Suzhou University. He was a Post-Doctoral Fellow with the Department of Computer Science, Concordia University, Montreal, QC, Canada, from September 2001 to August 2002, and a Visiting Assistant Professor with the Department of Computer Science, University of Vermont, Burlington, VT, USA, from September 2002 to February 2004. He was a Full Professor with the Department of Mathematics, Zhejiang University, Hangzhou, Zhejiang, China, from March 2004 to February 2013. He is currently a Full Professor with the School of Mathematical Sciences, Capital Normal University, Beijing, China. His research interests include combinatorics, coding theory, and information security and their interactions. He received the 2006 Hall Medal from the Institute of Combinatorics and its Applications. He is on the editorial board of Journal of Combinatorial Theory, Series A, IEEE Transactions on Information Theory, Designs, Codes and Cryptography, Journal of Combinatorial Designs, Journal of Algebraic Combinatorics, Science China Mathematics, and Applied Mathematics-A Journal of Chinese Universities.
\end{IEEEbiographynophoto}

\begin{IEEEbiographynophoto}{Ohad~Elishco}
(\emph{Member, IEEE}) received his B.Sc. degree in mathematics and his B.Sc., M.Sc., and Ph.D. degrees in electrical engineering from the Ben-Gurion University of the Negev, Israel, in 2012, 2013, and 2017, respectively. From 2017 to 2018, he held a post-doctoral position with the Department of Electrical Engineering, Massachusetts Institute of Technology. From 2018 to 2020, he held a post-doctoral position with the Department of Electrical Engineering, University of Maryland, College Park. He is currently an Assistant Professor at the School of Electrical and Computer Engineering, Ben-Gurion University of the Negev, Israel. His research interests include coding theory and dynamical systems.
\end{IEEEbiographynophoto}
\end{document}